\documentclass[10pt]{article}
\usepackage{latexsym}
\usepackage{amssymb}
\usepackage{amsmath}
\usepackage{graphicx}
\usepackage[latin1]{inputenc}

\newtheorem{theorem}{Theorem}[section]
\newtheorem{prop}{Proposition}[section]
\newtheorem{coro}{Corollary}[section]
\newtheorem{lemma}{Lemma}[section]
\newtheorem{remark}{Remark}[section]

\newcommand{\R}{\mathbb{R}}             
\newcommand{\N}{\mathbb{N}}             
\newcommand{\Z}{\mathbb{Z}}             %
\newcommand{\C}{\mathbb{C}}             
\renewcommand{\H}{\mathcal{H}}          
\newcommand{\B}{\mathcal{B}}            
\newcommand{\M}{\mathcal{M}}            

\renewcommand{\L}{L^{2}(\R)}            
\newcommand{\CO}{C_{0}^{\infty}(\R)}    
\newcommand{\comp}{C_{0}^{\infty}}
\newcommand{\COC}{C_{0}^{\infty}(\R;\C^4)}

\newcommand{\e}{\epsilon}
\newcommand{\half}{\frac{1}{2}}

\newcommand{\Ga}{\Gamma^1}         
\newcommand{\Gb}{\Gamma^2}         
\newcommand{\Gc}{\Gamma^3}         
\newcommand{\Gd}{\Gamma^0}         

\renewcommand{\l}{\langle}
\renewcommand{\r}{\rangle}
\newcommand{\x}{\l x \r}
\newcommand{\xsi}{\l \xi \r}

\newcommand{\F}{\mathcal{F}}           
\renewcommand{\d}{\partial}            

\newcommand{\Nu}{\mathcal{V}}          

\newcommand{\Section}[1]{\section{#1} \setcounter{equation}{0}}

\author{Thierry Daudé \footnote{Department of Mathematics and
    Statistics, McGill University, 805 Sherbrooke South West, Montréal
    QC, H3A 2K6} \, and François Nicoleau
    \footnote{Département de
    Mathématiques, Université de Nantes, 2, rue de la Houssinière, BP
    92208, 44322 Nantes Cedex 03}}
\title{Inverse Scattering in de Sitter-Reissner-Nordstr\"om black hole spacetimes}
\date{}

\begin{document}

\maketitle


\begin{abstract}
In this paper, we study the inverse scattering of massive charged Dirac fields in the exterior region
of (de Sitter)-Reissner-Nordstr\"om black holes. First we obtain a precise high-energy asymptotic expansion
of the diagonal elements of the scattering matrix (\textit{i.e.} of the transmission coefficients) and we
show that the leading terms of this expansion allows to recover uniquely the mass, the charge and the
cosmological constant of the black hole. Second, in the case of nonzero cosmological constant, we show
that the knowledge of the reflection coefficients of the scattering matrix on any interval of energy also permits to
recover uniquely these parameters.
\end{abstract}


\Section{Introduction}


This paper deals with inverse scattering problems in black hole spacetimes and is a continuation of our previous
work \cite{DN}. Here we shall study the inverse scattering of massive charged Dirac fields that propagate
in the outer region of (de Sitter)-Reissner-Nordstr\"om black holes, an important family of spherically symmetric,
charged exact solutions of the Einstein equations that will be thoroughly described in Section \ref{dS-RN}.
These spacetimes are completely characterized by three parameters: the mass $M>0$ and the electric charge
$Q \in \R$ of the black hole and the cosmological constant $\Lambda \geq 0$ of the universe. In what follows,
these parameters will be considered as the "unknowns" of our inverse problem. In fact, the inverse scattering problem
we have in mind is of the following type: we assume that we are observers living in the exterior region of a
(dS)-RN black hole, that is the region between the exterior event horizon of the black hole and the cosmological
horizon when $\Lambda > 0$, or the region lying beyond the exterior event horizon of the black hole when $\Lambda = 0$.
The geometry of the spacetime in which these observers live is thus fixed in some sense. But, what we don't assume
however is that these observers know the exact values of the parameters $M, Q$ and $ \Lambda$ "a priori". Hence the
natural question we adress is: do such observers have any means to measure or characterize uniquely these parameters
by an inverse scattering experiment?


Let us first describe more precisely the exact inverse scattering experiment studied in
this paper. A direct scattering theory for massive charged Dirac fields has been established
in \cite{Da} for RN black holes and more generally in \cite{Me} for dS-RN black holes. As shown in these papers,
the outer region of (dS)-RN black holes offers an original geometrical situation from the point of view of scattering
theory. These spacetimes possess indeed \emph{two distinct asymptotic regions}, namely either the exterior event horizon
of the black hole and the cosmological horizon when $\Lambda>0$, or the event horizon of the black hole and spacelike
infinity when $\Lambda=0$, which may have \emph{very different geometrical structures}. The first consequences concerning the propagation properties of Dirac fields are given in the following important result obtained in \cite{Da, Me}: the energy of Dirac fields contained in any compact set between the two asymptotic regions tends to zero when the time tends to infinity. Therefore, Dirac fields scatter toward these asymptotic regions at late times and moreover, they are shown to obey there simple but different equations. From the mathematical point of view, two distinct wave operators must be introduced according to the asymptotic region we consider. Let us denote for the moment the wave operators corresponding to the part of Dirac fields which scatters toward the event horizon of the black hole by $W^\pm_{(-\infty)}$ and the wave operators corresponding to the part of Dirac fields which scatters toward the cosmological horizon or spatial infinity by $W^\pm_{(+\infty)}$. These wave operators will be precisely defined in Section \ref{dS-RN}. Now the main result obtained in \cite{Da, Me} shows that the global wave
operators defined by
\begin{equation} \label{GlobalWO}
  W^\pm = W^\pm_{(-\infty)} + W^\pm_{(+\infty)},
\end{equation}
exist and are asymptotically complete. This permits to define a global scattering operator $S$ by the usual formula
$$
  S = (W^+)^* W^-.
$$

The scattering operator $S$ will be the main object of study of this paper. In fact, we rephrase and precise
our initial problem in the following way. We assume that our observers have access experimentally to the scattering
operator $S$. In particular, we assume that they may measure the expectation values of $S$, \textit{i.e.} they can
measure any quantities of the form $<S \, \psi, \phi>$ where $<.,.>$ denotes the scalar product of the energy Hilbert
space $\H$ on which $S$ acts and $\psi, \phi$ are any element of $\H$. The question we adress is now: is the knowledge
of $S$ and any of its related quantities a sufficient information to uniquely characterize the parameters $M, Q$ and
$\Lambda$ of (dS)-RN black holes?

We can in fact be more precise in the statement of the problem if we remark that the scattering operator $S$ can be
decomposed using (\ref{GlobalWO}) as
$$
  S = T_L + T_R + L + R,
$$
where
$$
  T_L = (W^+_{(+\infty)})^* W^-_{(-\infty)}, \quad T_R = (W^+_{(-\infty)})^* W^-_{(+\infty)},
$$
and
$$
   R = (W^+_{(+\infty)})^* W^-_{(+\infty)}, \quad L = (W^+_{(-\infty)})^* W^-_{(-\infty)}.
$$

Each of the terms in $S$ corresponds to a different inverse scattering experiment. For instance, the first two
terms $T_R$ and $T_L$ (in fact the diagonal elements of $S$) can be understood as \emph{transmission operators}.
Precisely, they correspond to the following experiment: a signal is emitted in the remote past from one
asymptotic region and is captured in the late future in the other asymptotic region. These two terms measure thus
the part of a signal which is transmitted from one asymptotic region to the other in a scattering process.
Conversely, the last two terms $L$ and $R$ (the anti-diagonal elements of $S$) can be understood as
\emph{reflection operators} and correspond to the opposite experiment: a signal is emitted in the remote past
from one asymptotic region and is captured in the late future in the same asymptotic region. Hence these two terms
measure the part of the signal which is reflected from an asymptotic region to itself in a scattering process.
Depending on the point of view of our observers, the quantities of interest will be thus the expectation values
$<T_R \psi, \phi>, <T_L \psi, \phi>$ and $<L \psi, \phi>, <R \psi, \phi>$ of the transmission and reflection
operators respectively. In this paper, we shall study in fact two types of inverse problems. Firstly, in the two
cases of RN black holes ($\Lambda =0$) and dS-RN black holes ($\Lambda>0$), we shall prove that the parameters
$M, Q, \Lambda$ are uniquely determined if we assume that the high energies of the transmission operators $T_R$
or $T_L$ are known. Secondly, in the case of dS-RN black holes only ($\Lambda > 0$), we shall prove the same
uniqueness result under the assumption that the reflection operators $L$ or $R$ are known on any (possibly small) interval of energy.


Let us now recall the results of \cite{DN} where the first kind of inverse problem was adressed in the case
of Reissner-Nordstr\"om black holes (\textit{i.e.} with only the two parameters $M,Q$ unknown and the cosmological
constant $\Lambda$ equal to $0$). Using the direct scattering theory for massless Dirac fields obtained in \cite{Da, N}
and a high energy asymptotic expansion of the expectation values $<T_R \psi, \phi>$ or $<T_L \psi, \phi>$
(as defined above), a partial answer was then given: the mass $M$ and the modulus of the charge $|Q|$ are uniquely
determined from the leading terms of this high energy asymptotic expansion. Note that the indecision of the sign of the
charge is not surprising in that case since the propagation of massless Dirac fields is only influenced by the geometry
of the black hole which in turn only depends on $|Q|$ (see the expression of the metric (\ref{F}) in Section \ref{dS-RN}). Moreover, it was mentioned in \cite{DN} (see also \cite{G} where a similar problem was studied) that the same method couldn't be applied to uniquely recover the parameters from the high energies of the reflection operators $R$ or $L$. The relevant quantities are in that case indeed non measurable. In this paper we continue our investigation and improve our results in several directions.


In Section \ref{L0}, we reconsider the case $\Lambda = 0$ corresponding to RN black holes but study the
inverse scattering of massive charged Dirac fields instead of massless Dirac fields. Using the same approach than in \cite{DN}, we show that the mass $M$ as well as the charge $Q$ are uniquely determined by the leading terms of the high energy asymptotic expansion of the transmission operators $T_R$ or $T_L$. In fact, the advantage of considering massive charged Dirac fields is that an explicit term associated to the interaction between the electric charge of the fields and that of the black hole appears in the equation and allows to recover $Q$ and not $|Q|$. From the mathematical side, the analysis turns out to be much more involved than in \cite{DN} because of two main reasons. First, \emph{massive} Dirac fields have completely distinct behaviours when approaching the different asymptotic regions. At the event horizon of the black hole for instance, the attraction exerced by the black hole is so strong that massive Dirac fields behave as massless Dirac fields. The dynamic there is very simple and will be shown to obey a system of transport equations along the null radial geodesics of the black hole. This is a consequence of the particular geometry (of hyperbolic type) near the event horizon (and more generally near any horizons). Conversely, RN black holes are asymptotically flat at spacelike infinity. There, the fields simply behave like massive Dirac fields in the usual Minkowski spacetime and the mass of the fields, slowing down the propagation, plays an important role. In consequence the dynamics near the two asymptotic regions are quite different and must be treated separatly. The second kind of difficulty comes from the appearance of long-range terms in the equation but only in one asymptotic region: spacelike infinity. This entails new technical difficulties such as a modification of the standard wave operators at infinity and we need to work harder to obtain the high energy asymptotic expansion of the transmission operators. We want to emphasize at last that the model studied in this part can be viewed as a good intermediate model before studying the same inverse problem in the more complicated geometrical setting of Kerr black holes. As shown in \cite{HaN} indeed, the appearance of long-range terms in the equation (even for massless Dirac fields) is compulsory in that case as a side effect of the rotation of the spacetime.

In Section \ref{LNot0}, we consider the case of nonzero cosmological constant $\Lambda > 0$, that is de Sitter-Reissner-Nordstr\"om black holes and we have three unknown parameters $M, Q, \Lambda$ a priori. The two asymptotic regions are the event horizon of the black hole and the cosmological horizon. Near these regions, massive Dirac fields behave as massless Dirac fields and as before, their propagation obeys essentially a system of transport equations along the null radial geodesics of the black hole. However, different oscillations appear in the dynamics near these two horizons, once again due to the interaction between the charge of the field and that of the black hole. In consequence, Dirac fields evolve according to slightly different dynamics in that case too. In Subsection \ref{dSHE}, using the results of the previous part, we shall obtain a high energy asymptotic expansion of the transmission operators $T_R$ and $T_L$ and again, we shall prove that the parameters $M, Q$ and $\Lambda$ are uniquely characterized by the leading terms of this asymptotic expansion. Then we consider inverse scattering experiments based on the knowledge of the reflection operators $R$ or $L$ on a (small) interval of energy. As already mentioned, a high energy aymptotic expansion of these reflection operators doesn't give any information and can't be used to solve the inverse problem. To study this case, we follow instead the usual stationary approach of inverse scattering theory on the line. We refer for instance to the review by Faddeev \cite{F} and to the important paper by Deift and Trubowitz \cite{DT} for a presentation of the method for Schrodinger operators and to the nice paper \cite{AKM} for a recent application to Dirac operators (see also \cite{G, HJKS}). In Subsection \ref{dSAE} we shall first obtain a stationary representation of the scattering operator $S$ in terms of the usual transmission and reflection "coefficients" (note that these turn out to be matrices in our case). This is done after a serie of simplifications of our model which happens finally to reduce to  a particular case of the model studied in \cite{AKM}. Then we use the analysis of \cite{AKM}, namely a classical Marchenko method based on a carefull study of the stationary solutions of the corresponding Dirac equation, to prove the following result: the knowledge of one of the reflection operators $L$ or $R$ at all energies is enough to uniquely characterize the parameters $M,Q$ and $\Lambda$. Eventually, we improve this result seeing that, in our model, the reflection operators $R$ or $L$ are in fact analytic in the energy variable on a small strip containing the real axis. Hence it is enough to know $R$ or $L$ on any interval of energy in order to uniquely know them for all energies. Applying the result of \cite{AKM}, this leads to the uniqueness of the parameters in that case too.


We finish this introduction saying a few words on the main technical tool used in Sections \ref{L0} and
\ref{LNot0} to prove our uniqueness results from the high energies of the transmission operators $T_R$ ot $T_L$.
These are based on a high-energy expansion of the scattering operator $S$ following an approach introduced by
Enss and Weder in \cite{EW1} in the case of multidimensional Schr\"odinger operators. (Note that the case of
multidimensional Dirac  operators in flat spacetime was treated later by Jung in \cite{Ju}). Their result can be
summarized as follows. Using purely time-dependent methods, they showed roughly speaking that the first term of
the high-energy expansion of $S$ is exactly the Radon transform of the potential they are looking for. Since
they work in dimension greater than two, this Radon transform can be inversed and the potential thus uniquely
recovered. In our problem however, due to the spherical symmetry of the black hole, we are led to study a family
of one dimensional Dirac equations and the above Radon transform simply becomes an integral of a one-dimensional
function, hence a number, and cannot be inversed. Fortunately in our models, it turns out that this integral can
be explicitely computed and gives in general already a physically relevant information. Nevertheless, it is not
enough to uniquely characterize all the parameters of the black hole. In fact, we need to calculate several
terms of the asymptotic (and thus obtain several integrals) to prove our result. To do this, we follow the
stationary technique introduced by one of us \cite{Ni1} which is close in spirit to the Isozaki-Kitada method
used in long-range scattering theory \cite{IK}. The basic idea is to replace the wave operators (and thus the
scattering operator) by explicit Fourier Integral Operators, called modifiers, from which we are able to compute
the high-energy expansion readily. The construction of these modifiers and the precise determination of their
phases and amplitudes will be given in a self-contained manner in Section \ref{L0}. Note also that the similar
results proved in our previous paper \cite{DN} couldn't be applied directly to our new model because of the
presence of long-range terms in the equation. At last we mention that, while this method was well-known for
Schr\"odinger operators and applied successfully to various situations (see \cite{Ar, Ni1, Ni2, Ni3}), it has
required some substantial modifications when applied to Dirac operators, essentially because of the
matrix-valued nature of the equation. To deal with these difficulties, we made an extensive use of the paper by
G\^atel and Yafaev \cite{GY} where a direct scattering theory of massive Dirac fields in flat spacetime was
studied and modifiers were constructed.


\Section{(De Sitter)-Reissner Nordstr\"om black holes and Dirac equation} \label{dS-RN}

In this section, we describe the geometry of the exterior regions of (de Sitter)-Reissner-Nordstr\"om black holes. In particular, we emphasize the different properties of the asymptotic regions mentioned in the introduction, clearly distinguishing between the cases of zero and nonzero cosmological constant $\Lambda$. We then express in a synthetic manner the equations that govern the evolution of massive charged Dirac fields in these spacetimes. We end up this section recalling the known direct scattering results of \cite{Da, Me} and introducing the scattering operator $S$.


\subsection{(De Sitter)-Reissner-Nordstr\"om black holes}

In Schwarzschild coordinates a (de Sitter)-Reissner-Nordstr\"om black hole is described by a four dimensional smooth manifold
$$
  \M = \mathbb{R}_{t} \times \R^+_r \times S_{\omega}^{2},
$$
equipped with the lorentzian metric
\begin{equation} \label{Metric}
  g = F(r)\,dt^{2} - F(r)^{-1} dr^{2} - r^{2}d\omega^{2},
\end{equation}
where
\begin{equation} \label{F}
  F(r) = 1-\frac{2M}{r}+\frac{Q^{2}}{r^{2}} - \frac{\Lambda r^2}{3},
\end{equation}
and $d\omega^2 = d\theta^{2}+\sin^{2}\theta \, d\varphi^{2}$ is the euclidean metric on the sphere $S^{2}$. The constants $M>0$, $Q \in \R$ appearing in (\ref{F}) are interpreted as the mass and the electric charge of the black hole and $\Lambda \geq 0$ is the cosmological constant of the universe. Observe that the function (\ref{F}) and thus the metric (\ref{Metric}) do not depend on the angular variables $\theta, \varphi \in S^2$ reflecting the fact that dS-RN black holes are spherically symmetric spacetimes.

The family $(\M, g)$ are in fact exact solutions of the Einstein-Maxwell equations
\begin{equation} \label{Einstein}
  G_{\mu \nu} = 8 \pi T_{\mu \nu}, \quad \quad G_{\mu \nu} = R_{\mu \nu} + \half R g_{\mu \nu} + \Lambda g_{\mu \nu},
\end{equation}
Here $G_{\mu \nu}, R_{\mu \nu}$ and $R$ denote respectively the Einstein tensor, the Ricci tensor and the scalar curvature of $(\M, g)$ while $T_{\mu \nu}$ is the energy-momentum tensor
\begin{equation} \label{EM-Tensor}
  T_{\mu \nu} = \frac{1}{4\pi} \big( F_{\mu \rho} F_\nu^{\ \rho} - \frac{1}{4} g_{\mu \nu} F_{\rho \sigma} F^{\rho \sigma} \big),
\end{equation}
where $F_{\mu \nu}$ is the electromagnetic two-form solution of the Maxwell equations $\nabla^\mu F_{\nu \rho} = 0$, $\nabla_{[\mu} F_{\nu \rho]} = 0$ and given here in terms of a global electromagnetic vector potential
\begin{equation} \label{EVPotential}
  F_{\mu \nu} = \nabla_{[\mu} A_{\nu]}, \quad A_\nu dx^\nu = -\frac{Q}{r} dt.
\end{equation}
We point out that any \emph{spherically symmetric} solutions of the Einstein equations (\ref{Einstein})-(\ref{EVPotential}) must belong (at least locally) to the family of dS-RN black holes defined by (\ref{Metric}) and (\ref{F}). This is a well-known uniqueness result due to Birkhoff (see for instance \cite{HE}). In particular, the results contained in this paper apply to this extended class of spacetimes.

The metric $g$ has two types of singularities. Firstly, the point $\{r = 0\}$ for which the function $F$ is singular. This is a
true singularity or \emph{curvature singularity} \footnote{It means that certain scalars obtained by contracting the Riemann tensor blow up when $r \to 0$.}. Secondly, the spheres whose radii are the roots of $F$ (note that the coefficient of the metric
$g$ involving $F^{-1}$ explodes in this case). We must distinguish here two cases. When the cosmological constant is positive $\Lambda>0$ and small enough, there are three positive roots $0 \leq r_- < r_0 < r_+ < +\infty$ . The spheres of radius $r_-, r_0$ and $r_+$ are called respectively Cauchy, event and cosmological horizons of the dS-RN black hole. When $\Lambda=0$, the number of these roots depends on the respective values of the constants $M$ and $Q$. In this paper we only consider the case $M > |Q|$ for which the function $F$ has two zeros at the values $r_- = M - \sqrt{M^2 - Q^2}$ and $r_0 = M + \sqrt{M^2-Q^2}$. The spheres of radius $r_-$ and $r_0$ are called respectively the Cauchy and event horizons of the RN black hole. 
In both situations, the horizons are not true singularities in the sense given for $\{r = 0\}$, but in fact \emph{coordinate singularities}. It turns out that, using appropriate coordinate systems, these horizons can be understood as regular null hypersurfaces that can be crossed one way but would require speeds greater than that of light to be crossed the other way. We refer to \cite{HE} and \cite{W} for a introduction to black hole spacetimes and their general properties.

As mentioned in the introduction, we shall consider in this paper inverse scattering experiments made by observers living in the exterior region of a (dS)-RN black hole, that is the region $\{r_0 < r < r_+\}$ when $\Lambda >0$ or the region $\{r_0<r<+\infty\}$ when $\Lambda=0$. It is thus important to understand the roles of the horizons as the natural boundaries of the exterior region. In Schwarzschild coordinates, it turns out that they are \emph{asymptotic regions} of spacetime. Precisely, this means that they are never reached in a finite time $t$ by incoming and outgoing null radial geodesics, \textit{i.e} the trajectories followed by classical light-rays aimed radially at the black hole and either at the cosmological horizon  if $\Lambda >0$ or at infinity if $\Lambda=0$. To see this point more easily, we introduce a new radial coordinate $x$,
called the Regge-Wheeler coordinate, which has the property of straightening the null radial geodesics and will, at the same time, greatly simplify the later analysis. Observing that for all $\Lambda \geq 0$ the function $F(r)$ in the metric (\ref{F}) remains always positive in the exterior region, it can be defined implicitely by the relation
\begin{equation} \label{RWImplicit}
  \frac{dr}{dx} = F(r) > 0,
\end{equation}
or explicitely, by
\begin{equation} \label{RWExplicit1}
  x = \frac{1}{2\kappa_0} \Big[ \log(r-r_0) - \int_{r_0}^r \big( \frac{1}{y-r_0} - \frac{2\kappa_0}{F(y)} \big) dy \Big] + C,
\end{equation}
where the quantity
$$
  \kappa_0 = \half F'(r_0) > 0,
$$
is called the surface gravity of the event horizon and $C$ is any constant of integration. Note that, when $\Lambda>0$, the Regge-Wheeler variable could be also defined explicitely by
\begin{equation} \label{RWExplicit2}
  x = \frac{1}{2\kappa_+} \Big[ \log(r_+-r) - \int_r^{r_+} \big( \frac{1}{r_+-y} + \frac{2\kappa_+}{F(y)} \big) dy \Big] + C,
\end{equation}
where the quantity
$$
  \kappa_+ = \half F'(r_+) < 0,
$$
is called the surfave gravity of the cosmological horizon. Moreover, in the case $\Lambda=0$, the expression (\ref{RWExplicit1}) simplifies as
\begin{equation} \label{RWExplicit3}
  x = r + \frac{1}{2\kappa_0} \log (r-r_0) + \frac{r_-^2}{r_0 - r_-} \log (r-r_-) + C.
\end{equation}

In the coordinate system $(t,x, \omega)$, it is easy to see from the logarithm in (\ref{RWExplicit1}) and (\ref{RWExplicit3}) and the positive sign of $\kappa_0$ that the event horizon $\{r=r_0\}$ is pushed away to $\{x = -\infty\}$ for all $\Lambda \geq 0$. Similarly it follows from (\ref{RWExplicit2}) and the negative sign of $\kappa_+$ that the cosmological horizon $\{r=r_+\}$ is pushed away to $\{x=+\infty\}$ when $\Lambda>0$. Hence in any case the Regge-Wheeler variable $x$ runs over the full real line $\R$. Moreover, by (\ref{RWImplicit}), the metric takes now the nice form
\begin{equation} \label{MetricRW}
  g = F(r) (dt^2 - dx^2) -r^2 d\omega^2,
\end{equation}
from which it is immediate to see that the incoming and outgoing null radial geodesics are generated by the vector fields $\frac{\partial}{\partial t} \pm \frac{\partial}{\partial x}$ and take the simple form
\begin{equation} \label{NullGeodesics}
  \gamma^{\pm}(t) = (t, x_0 \pm t, \omega_0), \ \ t \in   \R,
\end{equation}
where $(x_0, \omega_0) \in \R \times S^2$ are fixed. These are simply straight lines with velocity $\pm 1$ mimicking, at least in the $t-x$ plane, the situation of a one-dimensional Minkowski spacetime. At last, using (\ref{NullGeodesics}), we can check directly that the event horizon and the cosmological horizon (when $\Lambda>0$) are asymptotic regions of spacetime in the sense given above.

From now on we shall only consider the exterior region of dS-RN black holes and we shall work on the manifold
$\B = \R_t \times \Sigma$ with $\Sigma = \R_{x} \times S^2_{\omega}$, equipped with the metric (\ref{MetricRW}).
Such a manifold $\mathcal{B}$ is globally hyperbolic meaning that the foliation $\Sigma_t = \{t\} \times \Sigma$
by the level hypersurfaces of the function $t$, is a foliation of $\mathcal{B}$ by Cauchy hypersurfaces (see
\cite{W} for a definition of global hyperbolicity and Cauchy hypersurfaces). In consequence, we can view the
propagation of massive charged Dirac fields as an evolution equation in $t$ on the spacelike hypersurface
$\Sigma$, that is a cylindrical manifold having two distinct ends: $\{x=-\infty\}$ corresponding to the event
horizon of the black hole and $\{x=+\infty\}$ corresponding to the cosmological horizon when $\Lambda>0$ and to
spacelike infinity when $\Lambda=0$. Note that the geometries of these ends are distinct in general. The event
and cosmological horizons are indeed \emph{exponentially large ends} of $\Sigma$ whereas spacelike infinity is
an \emph{asymptotically flat end} of $\Sigma$ (in the latter, observe that the metric (\ref{F}) tends to the
Minkowski metric expressed in spherical coordinates when $r \to +\infty$). The difference between these
geometries will be easily seen from the distinct asymptotic behaviours of Dirac fields near these regions given
in the next subsection.


\subsection{Dirac equation and direct scattering results} \label{DiracEq}

Scattering theory for massive charged Dirac fields on the spacetime $\B$ has been the object of the papers \cite{Da, Me}. We briefly recall here the main results of these papers. In particular, we use the form of the Dirac equation obtained therein.

First, the evolution equation satisfied by massive charged Dirac fields in $\mathcal{B}$ can be written under the Hamiltonian form
\begin{equation} \label{DiracEquation}
  i \partial_{t} \psi = H \psi,
\end{equation}
where $\psi$ is a $4$-components spinor belonging to the Hilbert space
$$
  \H = L^2(\R \times S^2; \C^4),
$$
and the Hamiltonian $H$ is given by
\begin{equation} \label{FullDiracOperator}
  H =  \Ga D_x + a(x) D_{S^2} + b(x) \Gd + c(x).
\end{equation}
Here we use the following notations. The symbol $D_x$ stands for $-i\d_x$ whereas $D_{S^2}$ denotes the Dirac operator on $S^2$ which, in spherical coordinates, takes the form
\begin{equation} \label{DiracSphere}
  D_{S^2} = -i \Gb (\partial_{\theta} + \frac{\cot{\theta}}{2}) - \frac{i}{\sin{\theta}} \Gc   \partial_{\varphi}.
\end{equation}
The potentials $a, b, c$ are scalar smooth functions given in terms of the metric (\ref{Metric}) by
\begin{equation} \label{Potentials}
  a(x) = \frac{\sqrt{F(r)}}{r}, \quad b(x) = m \sqrt{F(r)}, \quad c(x) = \frac{qQ}{r},
\end{equation}
where $m$ and $q$ denote the mass and the electric charge of the fields respectively. Finally, the matrices $\Ga, \Gb, \Gc, \Gd$ appearing in (\ref{FullDiracOperator}) and (\ref{DiracSphere}) are usual $4 \times 4$ Dirac matrices that satisfy the anticommutation relations
\begin{equation} \label{AntiCom}
  \Gamma^i \Gamma^j + \Gamma^j \Gamma^i = 2 \delta_{ij} \textrm{Id}, \quad \forall i,j=0,..,3.
\end{equation}

Second, we use the spherical symmetry of the equation to simplify further the expression of the Hamiltonian $H$. Since, the Dirac operator $D_{S^2}$ has compact resolvent, it can be diagonalized into an infinite sum of matrix-valued multiplication operators. The eigenfunctions associated to $D_{S^2}$ are a generalization of the usual spherical harmonics called \emph{spin-weighted sphericl harmonics}. We refer to I.M. Gel'Fand and Z.Y. Sapiro \cite{GS} for a detailed presentation of
these generalized spherical harmonics and to \cite{Da, Me} for an application to our model. There exists thus a family of eigenfunctions $F_n^l$ of $D_{S^2}$ with the indexes $(l,n)$ running in the set $\mathcal{I} = \big\{ (l,n), l-|\half| \in \N, l- |n| \in \N \big\}$ which forms a Hilbert basis of $L^2(S^2; \C^4)$ with the following property. The Hilbert space $\H$ can then be decomposed into the infinite direct sum
\begin{displaymath}
  \H = \bigoplus_{(l,n) \in \mathcal{I}} \Big[L^{2}(\R_x; \C^4) \otimes F_{n}^{l} \Big] :=\bigoplus_{(l,n) \in \mathcal{I}} \H_{ln},
\end{displaymath}
where $\H_{ln} = L^{2}(\R_x; \C^4) \otimes F_{n}^{l}$ is identified with $L^2(\R; \C^4)$ and more important, we obtain the orthogonal decomposition for the Hamiltonian $H$
\begin{displaymath}
  H = \bigoplus_{(l,n) \in \mathcal{I}} H^{ln},
\end{displaymath}
with
\begin{equation} \label{DiracOperator}
  H^{ln} := H_{|\H_{ln}} = \Ga D_x + a_l(x) \Gb + b(x) \Gd + c(x),
\end{equation}
and $a_l(x) = - a(x) (l + \half)$. Note that the Dirac operator $D_{S^2}$ has been replaced in the expression of $H^{ln}$ by $-(l+\half) \Gb$ thanks to the good properties of the spin-weighted spherical harmonics $F_n^l$. The operator $H^{ln}$ is a selfadjoint operator on $\H_{ln}$ with domain $D(H^{ln}) = H^1(\R; \C^4)$. Finally we use the following representation for the Dirac matrices $\Ga, \Gb$ and $\Gd$ appearing in (\ref{DiracOperator})
\begin{equation} \label{DiracMatrices}
  \Ga = \left( \begin{array}{cccc} 1&0&0&0 \\0&1&0&0 \\ 0&0&-1&0 \\0&0&0&-1 \end{array} \right), \quad \Gb = \left( \begin{array}{cccc} 0&0&0&1 \\0&0&-1&0 \\ 0&-1&0&0 \\ 1&0&0&0 \end{array} \right), \quad \Gd = \left( \begin{array}{cccc} 0&0&-i&0 \\0&0&0&i \\ i&0&0&0 \\0&-i&0&0 \end{array} \right).
\end{equation}

In this paper it will be often enough to restrict our analysis to a fixed harmonic. To simplify notations we shall thus simply write $\H$, $H$ and $a(x)$ instead of $\H_{ln}$, $H^{ln}$ and $a_l(x)$ respectively and we shall indicate in the course of the text whether we work on the global problem or on a fixed harmonic.

Let us summarize now the direct scattering results obtained in \cite{Da, Me}. It is well known that the main information of interest in scattering theory concerns the nature of the spectrum of the Hamiltonian $H$. Our first result goes in this sense. Using essentially a Mourre theory (see \cite{Mo}), it was shown in \cite{Da, Me} that, for all $\Lambda \geq 0$,
$$
  \sigma_{\textrm{pp}}(H) = \emptyset, \quad \sigma_{\textrm{sing}}(H) = \emptyset.
$$
In other words, the spectrum of $H$ is purely absolutely continuous. In consequence, massive charged Dirac fields scatter toward the two asymptotic regions at late times and they are expected to obey simpler equations there. This is one of the main information encoded in the notion of \emph{wave operators} that we introduce now.

We first treat the case $\Lambda=0$ corresponding to RN black holes. From (\ref{F}) and (\ref{RWExplicit3}), the
potentials $a, b, c$ have very different asymptotics as $x \to \pm \infty$ (according to our discussion above
this reflects the fact that the geometries near the two asymptotic regions are very different). At the event
horizon, there exists $\alpha >0$ such that
\begin{equation} \label{P0H}
  |a(x)|, \ |b(x)|, \ |c(x) - c_0| = O (e^{\alpha x}), \quad x \to -\infty,
\end{equation}
where the constant $c_0$ is given by (see (\ref{Potentials}))
$$
  c_0 = \frac{qQ}{r_0}.
$$
Hence, the potentials $a, b, c-c_0$ are short-range when $x \to -\infty$ and we can choose the asymptotic dynamic generated by the Hamiltonian $H_0 = \Ga D_x + c_0$ as the comparison dynamic in this region. The Hamiltonian $H_0$ is a selfadjoint operator on $\H$ with its spectrum covering the full real line, \textit{i.e.} $\sigma(H_0) = \R$. Note finally that due to the simple diagonal form of the matrix $\Ga$, the comparison dynamic $e^{-itH_0}$ is essentially a system of transport equations along the curves $x \pm t$, that is the null radial geodesics of the black hole.

Conversely at infinity, the potentials $a, b, c$ have the asymptotics
\begin{equation} \label{P0I}
  |a(x)|, \ |b(x) - m|, \ |c(x)| = O(\frac{1}{x}), \quad x \to +\infty.
\end{equation}
The potentials $a, b-m, c$ are thus long-range potentials having Coulomb decay when $x \to +\infty$. The asymptotic dynamic there is generated by the Hamiltonian $H_0^m = \Ga D_x + m \Gd$, a classical one-dimensional Dirac Hamiltonian in Minkowski spacetime. The Hamiltonian $H_0^m$ is a selfadjoint operator on $\H$ and its spectrum has a gap, \textit{i.e.} $\sigma(H_0^m) = (-\infty, -m) \cup (+m, +\infty)$. Contrary to the preceding case, the asymptotic dynamic $e^{-itH_0^m}$ cannot be used alone as a comparison dynamic because of the long-range terms, but must be (Dollard)-modified.

In order to define this modification and for other use, we need to introduce the classical velocity operators
$$
  \Nu_0 = \Ga, \quad \Nu_m = D_x (H_0^m)^{-1},
$$
associated to the Hamiltonians $H_0$ and $H_0^m$ respectively. The classical velocity operators are selfadjoint operators on $\H$ and their spectra are simply $\sigma(\Ga) = \{-1,+1\}$ and $\sigma(\Nu_m) = [-1,+1]$. Let us also denote by $P_\pm$ and $P_\pm^m$ the projections onto the positive and negative spectrum of $\Ga$ and $\Nu_m$, \textit{i.e.}
$$
  P_\pm = \mathbf{1}_{\R^\pm}(\Ga), \quad P_\pm^m = \mathbf{1}_{\R^\pm}(\Nu_m).
$$
As shown in \cite{Da}, a great interest of these projections is that they permit to separate easily the part of the fields that propagate toward the event horizon and the part of the fields that propagate toward infinity. They will be used in the definition of the wave operators below. Moreover, the classical velocity operator $\Nu_m$ enters in the expression of the Dollard modified comparison dynamic at infinity proposed in \cite{Da} and given by
\begin{equation} \label{Dollard}
  U(t) = e^{-it H_0^m} e^{-i \int_0^t \big[ (b(s\Nu_m)-m) m (H_0^m)^{-1} + c(s\Nu_m) \big] ds}.
\end{equation}
Let us make here two comments. First, the potential $a(x) \Gb$ turns out to be a "false" long-range term. This is clear from (\ref{Dollard}) where the asymptotic dynamic $e^{-itH_0^m}$ has been modified by an extra phase which only involves the long-range potentials $b$ and $c$. We refer to \cite{Da} for an explanation of this particular point. Second, we shall propose in Section \ref{L0} a new \emph{time-independent} modification of the comparison dynamic $e^{-itH_0^m}$ which will be a direct byproduct of our construction of modifiers in the spirit of Isozaki-Kitada's work \cite{IK}. This new modification will be shown to be equivalent to the Dollard modification (\ref{Dollard}) in Theorem \ref{Isozaki-Kitada}.

We are now in position to introduce the wave operators associated to $H$. At the event horizon, we define
\begin{equation} \label{WOH1}
  W^\pm_{(-\infty)} = s-\lim_{t \to \pm \infty} e^{itH} e^{-itH_0} P_\mp,
\end{equation}
whereas at infinity, we define
\begin{equation} \label{WOI1}
  W^\pm_{(+\infty)} = s-\lim_{t \to \pm \infty} e^{itH} U(t) P^m_\pm.
\end{equation}
Finally, the global wave operators are given by
\begin{equation} \label{Global1}
  W^\pm = W^\pm_{(-\infty)} + W^\pm_{(+\infty)}
\end{equation}
Note here our use of the projections $P_\pm$ and $P_\pm^m$ to separate the part of the field propagating toward the event horizon to the part of the field propagating toward infinity. In fact without these projections, the wave operators (\ref{WOH1}) and (\ref{WOI1}) wouln't exist at all. More precisely the main result of \cite{Da} is
\begin{theorem} \label{WO+}
  The wave operators $W^\pm_{(-\infty)}$, $W^\pm_{(+\infty)}$ and $W^\pm$ exist on $\H$. Moreover, the global wave operators $W^\pm$ are partial isometries with initial spaces $\H^\pm_{\textrm{scat}} = P_\mp(\H) + P_\pm^m(\H)$ and final space $\H$. In particular, $W^\pm$ are asymptotically complete, \textit{i.e.} Ran $W^\pm = \H$.
\end{theorem}
As a direct consequence of Theorem \ref{WO+}, we can define the scattering operator $S$ by the usual formula
\begin{equation} \label{ScatteringOp}
  S = (W^+)^* W^-.
\end{equation}
It is clear that $S$ is a well-defined operator on $\H$ and a partial isometry from $\H^-_{\textrm{scat}}$ into $\H^+_{\textrm{scat}}$.

We now treat the case $\Lambda>0$ corresponding to dS-RN black holes wich turns out to be a little bit more symmetric at the two (event and cosmological) horizons. According to (\ref{F}), (\ref{RWExplicit1}) and (\ref{RWExplicit2}), the potentials $a, b, c$ have the following asymptotics as $x \to \pm \infty$. There exists $\alpha > 0$ such that
\begin{equation} \label{Pab}
  |a(x)|, \ |b(x)| = O (e^{-\alpha |x|}), \quad |x| \to \infty,
\end{equation}
and
\begin{eqnarray}
  |c(x) - c_0| & = & O (e^{\alpha x}), \quad x \to -\infty, \label{Pc1} \\
  |c(x) - c_+| & = & O (e^{-\alpha x}), \quad   x \to +\infty, \label{Pc2}
\end{eqnarray}
where the constants $c_0$ and $c_+$ are given by (see (\ref{Potentials}))
\begin{equation} \label{C0C+}
  c_0 = \frac{qQ}{r_0}, \quad c_+ = \frac{qQ}{r_+}.
\end{equation}
Hence, the potentials $a, b$ are short-range when $x \to \pm \infty$ and $c-c_0$ and $c-c_+$ are short-range when $x \to -\infty$ and $x \to +\infty$ respectively. At the event horizon, we choose as before the asymptotic dynamic generated by the Hamiltonian $H_0 = \Ga D_x + c_0$ as the comparison dynamic while, at the cosmological horizon, we choose the asymptotic dynamic generated by the Hamiltonian $H_+ = \Ga D_x + c_+$ as the comparison dynamic. The Hamiltonians $H_0$ and $H_+$ are clearly selfadjoint operators on $\H$ and their spectra are exactly the real line, \textit{i.e.} $\sigma(H_0) = \sigma(H_+) = \R$. We observe eventually that the dynamics $e^{-itH_0}$ and $e^{-itH_+}$ are essentially a system of transport equations along the null radial geodesics of the black hole but they differ by the distinct oscillations $e^{-itc_0}$ and $e^{-itc_+}$.

We need the classical velocity operators associated to $H_0$ and $H_+$ in order to separate the part of the fields that propagate toward the event horizon and the part of the fields that propagate toward the cosmological horizon. It turns out that they are equal to $\Nu_0 = \Ga$ in both cases and the associated projections onto the positive and negative spectrum are still $P_\pm$. Thus we can introduce the wave operators as before. At the event horizon, we define
\begin{equation} \label{WOH2}
  W^\pm_{(-\infty)} = s-\lim_{t \to \pm \infty} e^{itH} e^{-itH_0} P_\mp,
\end{equation}
and at the cosmological horizon, we define
\begin{equation} \label{WOI2}
  W^\pm_{(+\infty)} = s-\lim_{t \to \pm \infty} e^{itH} e^{-itH_+} P_\pm.
\end{equation}
Finally, the global wave operators are given by
\begin{equation} \label{Global2}
  W^\pm = W^\pm_{(-\infty)} + W^\pm_{(+\infty)}
\end{equation}
The main result of \cite{Me} is
\begin{theorem} \label{WO-}
  The wave operators $W^\pm_{(-\infty)}$, $W^\pm_{(+\infty)}$ and $W^\pm$ exist on $\H$. Moreover, the global wave operators $W^\pm$ are isometries on $\H$. In particular, $W^\pm$ are asymptotically complete, \textit{i.e.} Ran $W^\pm = \H$.
\end{theorem}
Thanks to Theorem \ref{WO-}, we can define the scattering operator $S$ as in (\ref{ScatteringOp}) by $S = (W^+)^* W^-$ which is a well-defined isometry on $\H$.

We deduce from the previous discussion that, for all $\Lambda \geq 0$, the scattering operator $S$ is a well-defined operator on $\H$. For all $\psi, \phi \in \H$, we shall consider in the following the expectation values of $S$, given by
$<S \psi, \phi>$, as the known data of our inverse problem. Moreover, using (\ref{Global1}) and (\ref{Global2}), we observe that these expectation values can be decomposed into $4$ natural components
$$
  <S \psi, \phi> \ = \ <W^- \psi, W^+ \phi>\ = \ <T_R \psi, \phi> + <T_L \psi, \phi> + <L \psi, \phi> + <R \psi, \phi>,
$$
where
\begin{eqnarray}
  <T_R \psi, \phi>  = <W^-_{(+\infty)} \psi, W^+_{(-\infty)} \phi>, & <T_L \psi, \phi> = <W^-_{(-\infty)} \psi,  W^+_{(+\infty)} \phi>, \label{T} \\
  <L \psi, \phi>  =  <W^-_{(-\infty)} \psi, W^+_{(-\infty)} \phi>, & <R \psi, \phi> = <W^-_{(+\infty)} \psi, W^+_{(+\infty)} \phi>. \label{R}
\end{eqnarray}
It follows from our definitions of the wave operators (\ref{WOH1}), (\ref{WOH2}) and (\ref{WOI1}), (\ref{WOI2}) that the previous qantities can be interpreted in terms of transmission and reflection between the different asymptotic regions, \textit{i.e.} $\{x=-\infty\}$ for the event horizon of the black hole and $\{x=+\infty\}$ for either spacelike infinity if $\Lambda=0$, or the cosmological horizon if $\Lambda>0$. For instance, $<T_R \psi, \phi>$ corresponds to the part of a signal transmitted from $\{x=+\infty\}$ to $\{x=-\infty\}$ in a scattering process whereas the term $<T_L \psi, \phi>$ corresponds to the part of a signal transmitted from $\{x=-\infty\}$ to $\{x=+\infty\}$. Hence $T_R$ stands for "transmitted from the right" and $T_L$ for "transmitted from the left". Conversely, $<L \psi, \phi>$ corresponds to the part of a signal reflected from $\{x=-\infty\}$ to $\{x=-\infty\}$ in a scattering process whereas the term $<R \psi, \phi>$ corresponds to the part of a signal reflected from $\{x=+\infty\}$ to $\{x=+\infty\}$.


\Section{The inverse problem when $\Lambda=0$} \label{L0}

In this section, we study the inverse problem at high energy in the case $\Lambda = 0$ that corresponds to RN black holes. Let us recall here that all the results and formulae given hereafter are always obtained on a fixed spin-weighted spherical harmonic. Therefore the notations $\H, H, a(x)$ are a shorthand for $\H_{ln}, H^{ln}, a_l(x)$ defined in the preceding Section. In order to state our main result, we make two assumptions. \\

\noindent\emph{Assumption 1}: We assume that our observers may measure the high energies of the transmitted operators $T_R$ or $T_L$. Precisely, we assume that one of the following functions of $\lambda \in \R$
$$
  F_l(\lambda) = <T_R e^{i\lambda x} \psi, e^{i\lambda x} \phi>, \quad \quad G_l(\lambda) = <T_L e^{i\lambda x} \psi, e^{i\lambda x} \phi>,
$$
are known for all large values of $\lambda$, for all $l \in \N$ where $l$ indexes the spin-weighted spherical
harmonics and for all $\psi, \phi \in \H$ with $\psi, \phi \in \comp(\R; \C^4)$. \\

\noindent
\emph{Assumption 2}: We also assume that the mass $m$ and the charge $q$ of the Dirac fields considered in these inverse scattering experiments are known and fixed. Moreover we assume that $q \ne 0$ since the case $q=0$ is similar to the one treated \cite{DN}.  \\

The main result of this section is now summarized in the following Theorem
\begin{theorem} \label{MainThm1}
  Under assumptions 1 and 2, the parameters $M$ and $Q$ of the RN black hole are uniquely determined.
\end{theorem}

Following our previous paper \cite{DN}, the proof of Theorem \ref{MainThm1} will be based on a high-energy asymptotic expansion of the functions $F_l(\lambda)$ and $G_l(\lambda)$ when $\lambda \to + \infty$. Precisely we shall prove the following formulae:
\begin{theorem}[Reconstruction formulae] \label{ReconstructionFormula}
  Let $\psi, \phi \in \comp(\R ; \C^4)$. Then for $\lambda$ large, we obtain
  \begin{eqnarray}
    F_l(\lambda) & = & <\Theta(x) P_- \psi, P_- \phi> + \frac{i}{2\lambda} < \mathcal{A}(x) P_- \psi, P_- \phi> + \ O(\lambda^{-2}), \label{RF1} \\
    G_l(\lambda) & = & <\Theta(x) P_+ \psi, P_+ \phi> - \frac{i}{2\lambda} < \mathcal{A}(x) P_+ \psi, P_+ \phi> + \ O(\lambda^{-2}), \label{RF2}
  \end{eqnarray}
  where $\theta(x)$ and $\mathcal{A}(x)$ are multiplication operators given by
  \begin{equation} \label{Theta}
    \Theta(x) = e^{-i\int_{-\infty}^0 [c(s)-c_0]ds + i c_0 x},
  \end{equation}
  \begin{equation}\label{A(x)}
\quad \mathcal{A}(x) = \Theta(x) \Big(  \int_{-\infty}^{+\infty} a_l^2(s) ds + \int_{-\infty}^0 b^2(s) ds
+  \int_0^{+\infty} (b(s)-m)^2 ds +  m^2 x \Big).
  \end{equation}
\end{theorem}
\begin{remark}
  In Theorem \ref{ReconstructionFormula}, we have once again emphasized the dependence of the functions               $F_l(\lambda)$ and $ G_l(\lambda)$ on the parameter $l$ since the reconstruction formulae (\ref{RF1}) and (\ref{RF2}) can be    derived if we work on a fixed spin-weighted spherical harmonic only. Nevertheless, as indicated in \emph{Assumption 1} we      shall need to know these formulae on all spin-weighted spherical harmonics, hence for all $l \in \N$, in order to prove the     uniqueness result stated in Theorem \ref{MainThm1}.
\end{remark}


\begin{remark} \label{Important0}
  In the reconstruction formulae of Thm \ref{ReconstructionFormula}, the constant terms $\int_{-\infty}^0 [c(s)-c_0]ds $
  in (\ref{Theta}) and $\int_{-\infty}^0 b^2(s) ds +  \int_0^{+\infty} (b(s)-m)^2 ds$ in (\ref{A(x)}),
  that may appear unnatural at first sight since they depend explicitely on the particular value $0$ of the
  Regge-Wheeler variable $x$, are in fact due to our particular choice of Dollard modification in the definition of
  the modified wave operators $W^\pm_{(+\infty)}$. Recall here indeed that there is no \emph{canonical} choice for
  the (necessary) modifications entailed by the presence of long-range potentials at infinity. This point can be
  clearly seen for instance from the Isozaki-Kitada modifications -constructed in the next
  Subsection- whose phases are defined only up to a constant of
  integration (see (\ref{Phase}) and Remark \ref{Important1} after it).
  We emphasize moreover that this constant of integration are not physically measurable
  and we can check indeed that they do not play any role in our proof of the uniqueness of the parameters.

\end{remark}

We now explain our strategy to prove Theorem \ref{ReconstructionFormula}. Using (\ref{WOH1}), (\ref{WOI1}), (\ref{T}) and the fact that $e^{i\lambda x}$ corresponds to a translation by $\lambda$ in momentum space, we first rewrite $F_l(\lambda)$ and  $G_l(\lambda)$ as follows
\begin{eqnarray}
  F_l(\lambda) & = & <W_{(+\infty)}^-(\lambda) \psi, W_{(-\infty)}^+(\lambda) \phi>, \label{ExpF} \\
  G_l(\lambda) & = & <W_{(-\infty)}^-(\lambda) \psi, W_{(+\infty)}^+(\lambda) \phi>, \label{ExpG}
\end{eqnarray}
with
\begin{eqnarray*}
  W_{(-\infty)}^\pm(\lambda) & = & e^{-i\lambda x} W_{(-\infty)}^\pm e^{i\lambda x} =  \ s-\lim_{t \to \pm \infty} e^{itH(\lambda)} e^{-itH_0(\lambda)} P_\mp,  \\
  W_{(+\infty)}^\pm(\lambda) & = & e^{-i\lambda x} W_{(+\infty)}^\pm e^{i\lambda x} =  \ s-\lim_{t \to \pm \infty} e^{itH(\lambda)} e^{-i X(t,\lambda)} e^{-it H_0^m(\lambda)} P_\pm^{m,\lambda},
\end{eqnarray*}
where we use the notations
$$
  H(\lambda) = \Ga (D_x + \lambda) + a(x) \Gb + b(x) \Gd + c(x), \quad H_0(\lambda) = \Ga (D_x + \lambda) + c_0,
$$
$$
  H_0^m(\lambda) = \Ga(D_x + \lambda) + m \Gd, \quad \Nu_m(\lambda) = (D_x+\lambda) \big(H_0^m(\lambda)\big)^{-1}, \quad P_\pm^{m,\lambda} = \mathbf{1}_{\R^\pm}(\Nu_m(\lambda)),
$$
$$
  X(t,\lambda) = \int_0^t \big[ (b(s\Nu_m(\lambda))-m) m (H_0^m(\lambda))^{-1} + c(s\Nu_m(\lambda)) \big] ds.
$$
In order to obtain an asymptotic expansion of the functions $F_l(\lambda)$ and $G_l(\lambda)$, it is thus enough to obtain an asymptotic expansion of the $\lambda$-shifted wave operators $W_{(\pm\infty)}^\pm(\lambda)$. To do this, we follow the procedure exposed in \cite{Ni1, Ni2}, procedure inspired by the well-known Isozaki-Kitada method \cite{IK} developed in the setting of long-range stationary scattering theory. It consists simply in replacing the wave operators $W_{(\pm\infty)}^\pm(\lambda)$ by ``well-chosen'' energy modifiers $J_{(\pm\infty)}^\pm(\lambda)$, defined as Fourier Integral Operators (FIO) with explicit phases and amplitudes. Well-chosen here means practically that we look for $J_{(\pm\infty)}^\pm(\lambda)$ satisfying for $\lambda$ large enough
\begin{eqnarray}
  W_{(-\infty)}^\pm(\lambda) \psi & = & \lim_{t \to \pm \infty} e^{itH(\lambda)} J_{(-\infty)}^\pm(\lambda) e^{-itH_0(\lambda)} P_\mp \psi, \label{Property1} \\
  W_{(+\infty)}^\pm(\lambda) \psi & = & \lim_{t \to \pm \infty} e^{itH(\lambda)} J_{(+\infty)}^\pm(\lambda) e^{-itH^m_0(\lambda)} P_\pm^{m,\lambda} \psi, \label{Property2}
\end{eqnarray}
and
\begin{equation} \label{Property3}
  \| (W_{(\pm\infty)}^\pm(\lambda) - J_{(\pm\infty)}^\pm(\lambda)) \psi \| =  O(\lambda^{-2}),
\end{equation}
for any fixed $\psi \in \H$ such that $\psi \in \comp(\R ; \C^4)$. Note that the decay $O(\lambda^{-2})$ in (\ref{Property3}) could be improved to any inverse power decay but turns out to be enough to our purpose here. In particular if we manage to construct such $J_{(\pm\infty)}^\pm(\lambda)$ satisfying (\ref{Property3}) then we obtain by (\ref{ExpF}) and (\ref{ExpG})
\begin{eqnarray}
  F_l(\lambda) & = & <J_{(+\infty)}^-(\lambda) \psi, J_{(-\infty)}^+(\lambda) \psi> + \ O(\lambda^{-2}), \label{Mod} \\
  G_l(\lambda) & = & <J_{(-\infty)}^-(\lambda) \psi, J_{(+\infty)}^+(\lambda) \psi> + \ O(\lambda^{-2}), \nonumber
\end{eqnarray}
from which we can calculate the first terms of the asymptotics easily.

Let us here give a simple but useful result which allows us to simplify slightly the expressions of (\ref{Property1}) and (\ref{Property2}).
\begin{lemma} \label{Scalar}
  For all $\xi \in \R^*$, set
  \begin{equation} \label{nu}
    \nu^\pm(\xi) = \pm \textrm{sgn}(\xi) \sqrt{\xi^2 + m^2}.
  \end{equation}
  Then, for all $\psi$ with $supp \  {\hat{\psi}} \subset \R^*$ ,
  \begin{equation} \label{Scalar1}
    e^{-itH_0^m} P^m_{\pm} \psi = e^{-it \nu^\pm(D_x)} P^m_{\pm} \psi.
  \end{equation}
  Moreover,
  \begin{equation} \label{Scalar2}
    e^{-itH_0} P_\pm  = e^{\mp it D_x -it c_0} P_\pm .
  \end{equation}
\end{lemma}
\textit{Proof}: The Fourier representation of the operator $H_0^m$ is $\Ga \xi + m \Gd$ and has precisely one positive eigenvalue $\sqrt{\xi^2 + m^2}$ and one negative eigenvalue $-\sqrt{\xi^2 + m^2}$. Similarly, the Fourier representation of the classical velocity operator $\Nu_m$ is $\frac{\xi}{\xi^2 + m^2} (\Ga \xi + m \Gd)$. Hence, for $\xi >0$, $P^m_+$ is the projection onto the positive spectrum of $\Ga \xi + m \Gd$ and $P^m_-$ is the projection onto the negative spectrum of $\Ga \xi + m \Gd$. For $\xi<0$, it is the opposite. This implies immediately (\ref{Scalar1}). Finally the equality (\ref{Scalar2}) is a direct consequence of the definitions of $H_0$ and $P_\pm$. $\diamondsuit$

According to Lemma \ref{Scalar}, the projections $P_\pm$ and $P^m_\pm$ allow us to "scalarize" the Hamiltonians $H_0$ and $H_0^m$ in the expressions (\ref{Property1}) and (\ref{Property2}) of $W_{(\pm\infty)}^\pm(\lambda)$. Precisely these expressions read now
\begin{eqnarray}
  W_{(-\infty)}^\pm(\lambda) \psi & = & \lim_{t \to \pm \infty} e^{itH(\lambda)} J_{(-\infty)}^\pm(\lambda) e^{\mp it (D_x+\lambda) -itc_0} P_\mp \psi, \label{Prop1} \\
  W_{(+\infty)}^\pm(\lambda) \psi & = & \lim_{t \to \pm \infty} e^{itH(\lambda)} J_{(+\infty)}^\pm(\lambda) e^{-it \nu^\pm(D_x+\lambda)} P_\pm^{m,\lambda} \psi. \label{Prop2}
\end{eqnarray}
This minor simplification will be important in the forthcoming construction of the modifiers $J_{(\pm\infty)}^\pm(\lambda)$

Before entering into the details, let us give a hint on how to construct the modifiers $J_{(\pm\infty)}^\pm(\lambda)$ a priori defined as FIOs with ``scalar'' phases $\varphi_{(\pm\infty)}^\pm(x,\xi,\lambda)$ and ``matrix-valued'' amplitudes $p_{(\pm\infty)}^\pm(x,\xi,\lambda)$, \textit{i.e.} defined for all $\psi \in \H$ by
$$
  J_{(\pm\infty)}^\pm(\lambda) \psi = \frac{1}{\sqrt{2\pi}} \int_\R e^{i\varphi_{(\pm\infty)}^\pm(x,\xi,\lambda)}  p_{(\pm\infty)}^\pm(x,\xi,\lambda) \hat{\psi}(\xi) d\xi.
$$
If we assume for instance that (\ref{Prop2}) is true then we easily get
\begin{equation} \label{WJ}
  (W_{(+\infty)}^\pm(\lambda) - J_{(+\infty)}^\pm(\lambda)) \psi = i \int_0^{\pm \infty} e^{itH(\lambda)} C_{(+\infty)}^\pm(\lambda) e^{-it \nu^\pm(D_x+\lambda)} P_\pm^{m,\lambda} \psi dt,
\end{equation}
where
\begin{equation} \label{C}
  C_{(+\infty)}^\pm(\lambda) := H(\lambda) J_{(+\infty)}^\pm(\lambda) - J_{(+\infty)}^\pm(\lambda) \nu^\pm(D_x+\lambda),
\end{equation}
are also FIOs with phases $\varphi_{(+\infty)}^\pm(x,\xi,\lambda)$ and amplitudes $c_{(+\infty)}^\pm(x,\xi,\lambda)$. From (\ref{WJ}) we get the simple estimate
\begin{equation} \label{Property4}
  \| (W_{(+\infty)}^\pm(\lambda) - J_{(+\infty)}^\pm(\lambda)) \psi \| \leq \int_0^{\pm \infty} \| C_{(+\infty)}^\pm(\lambda) \ e^{-it \nu^\pm(D_x+\lambda)} P_\pm^{m,\lambda} \psi  \| dt.
\end{equation}
In order that (\ref{Property3}) be true it is then clear from (\ref{Property4}) that the FIOs
$C_{(+\infty)}^\pm(\lambda)$ have to be ``small'' in some sense. Precisely we shall need that the
amplitudes $c_{(+\infty)}^\pm(x,\xi,\lambda)$ be short-range in the variable $x$ at infinity (\textit{i.e.}
when $x \to +\infty$) and of order $O(\lambda^{-2})$ when $\lambda \to +\infty$. Note here the role played by
the projections $P_\pm^{m,\lambda}$ which allow us to consider the part of the Dirac fields that propagate
toward infinity. This explains why the amplitudes $c_{(+\infty)}^\pm(x,\xi,\lambda)$
must short-range in the variable $x$ \emph{only at infinity}.
Similarly, for the construction of the modifiers  $J_{(-\infty)}^\pm(\lambda)$, we shall require that the amplitudes
$c^\pm_{(-\infty)}(x,\xi,\lambda)$ of the corresponding operators $C^\pm_{(-\infty)}(\lambda)$ be short-range in the
variable $x$ only at the event horizon (\textit{i.e.} when $x \to -\infty$) and of order $O(\lambda^{-2})$ when
$\lambda \to +\infty$.


\subsection{Asymptotics of $W_{(+\infty)}^\pm(\lambda)$} \label{AsympWI}

In this subsection, we construct the modifiers $J_{(+\infty)}^\pm(\lambda)$ and give the asymptotics of $W_{(+\infty)}^\pm(\lambda)$ when $\lambda \to +\infty$. For simplicity, we shall omit the lower index $(+\infty)$ in all the objects defined hereafter.

We first look at the problem at fixed energy (\emph{i.e.} we take $\lambda = 0$ in the previous formulae). Hence we aim to construct modifiers $J^\pm$ with scalar phases $\varphi^\pm(x,\xi)$ and matrix-valued amplitudes $p^\pm(x,\xi)$ such that the amplitudes $c^\pm(x,\xi)$ of the operators $C^\pm = H J^\pm - J^\pm \nu^\pm(D_x)$ be short-range in $x$ when $x \to +\infty$. We adapt here to our case the treatment given by Gâtel and Yafaev in \cite{GY} where a similar problem was considered in Minkowski spacetime (see also our recent paper \cite{DN}).

The operators $C^\pm$ are clearly FIOs with phases $\varphi^\pm(x,\xi)$ and amplitudes
\begin{equation} \label{AmplitudeC}
  c^\pm(x,\xi) = B^\pm(x,\xi) p^\pm(x,\xi) - i\Ga \d_x p^\pm(x,\xi),
\end{equation}
where
\begin{equation} \label{B}
  B^\pm(x,\xi) = \Ga \d_x \varphi^\pm(x,\xi) + a(x) \Gb + b(x) \Gd + c(x) - \nu^\pm(\xi).
\end{equation}
As usual we look for phases $\varphi^\pm$ close to $x \xi$ and amplitudes $p^\pm$ close to $1$. So the term $\d_x p^\pm$ in (\ref{AmplitudeC}) should be short-range et can be neglected in a first approximation. With $p^\pm=1$, we are thus led to solve $B^\pm =0$. However a direct calculation leads then to matrix-valued phases $\varphi^\pm$ whereas we look for scalar ones. We follow \cite{GY} and solve in fact $(B^\pm)^2 = 0$. Using crucially the anticommutation properties of the Dirac matrices (\ref{AntiCom}), we get the new equation
\begin{equation} \label{Ba1}
  (B^\pm)^2 = (\d_x \varphi^\pm)^2 + a^2 + b^2 +(c-\nu^\pm)^2 + 2(c-\nu^\pm) (B^\pm -c +\nu^\pm) = 0.
\end{equation}
If we put $B^\pm = 0$ in (\ref{Ba1}), we obtain the scalar equation
\begin{equation} \label{Ba2}
  r^\pm(x,\xi) := (\d_x \varphi^\pm)^2 + a^2 + b^2 - (c-\nu^\pm)^2 = 0.
\end{equation}
We look for an approximate solution of (\ref{Ba2}) of the form $\varphi^\pm(x,\xi) = x \xi + \phi^\pm(x,\xi)$
where $\phi^\pm(x,\xi)$ should be a priori relatively small in the variable $x$. Recalling that $(\nu^\pm)^2 =
\xi^2 + m^2$ by (\ref{nu}), we must then solve
\begin{equation} \label{Ba3}
  2\xi \d_x \phi^\pm + (\d_x \phi^\pm)^2 + a^2 + (b^2-m^2) - c^2 +2 c \nu^\pm = 0.
\end{equation}
If we neglect $(\d_x\phi^\pm)^2$ in (\ref{Ba3}), we finally get
\begin{equation} \label{Ba4}
  2 \xi \d_x \phi^\pm = -\big[a^2 + (b^2-m^2) - c^2 + d^\pm \big],
\end{equation}
where we have introduced the notation $d^\pm(x,\xi) = 2 c(x) \nu^\pm(\xi)$. Note that by (\ref{P0I}) and
(\ref{nu}), the following estimate holds
\begin{equation} \label{d}
  \forall \alpha, \beta \in \N, \quad |\d_x^\alpha \d_\xi^\beta d^\pm(x,\xi)| \leq C_{\alpha \beta} \x^{-1-\alpha} \xsi^{1-\beta}, \quad \forall x \in \R^+, \ \forall \xi \in \R^*.
\end{equation}


Therefore, using (\ref{P0I}) again and the previous estimate (\ref{d}), we see that $a^2 -c^2$ is short-range
when $x \to +\infty$ whereas $b^2-m^2$ and $d^\pm$ are long-range (of Coulomb type) when $x \to +\infty$.
Hence we can define two solutions of (\ref{Ba4}) for all $\xi \ne 0$ as follows
\begin{equation} \label{Phase}
  \phi^\pm(x,\xi) = \frac{1}{2\xi} \int_x^{+\infty} [a^2(s) - c^2(s)] ds -\frac{1}{2\xi}
  \int_0^x \big[(b^2(s) - m^2) + d^\pm(s,\xi) \big] ds + \frac{1}{2\xi} \int_0^{+\infty} (b(s)-m)^2 ds .
\end{equation}

\begin{remark} \label{Important1}
  Let us emphasize that we only add the quantity $\frac{1}{2\xi} \int_0^{+\infty} (b(s)-m)^2 ds$
  in (\ref{Phase}) in order to prove that the Isozaki-Kitada and the Dollard modifications coincide
  (see Theorem \ref{Isozaki-Kitada}). In the general case however, the phases $\tilde{\phi}^\pm(x,\xi)$,
  solutions of (\ref{Ba4}) would clearly take the form for all $\xi \ne 0$
\begin{equation} \label{NewPhase}
  \tilde{\phi}^\pm(x,\xi) = \frac{1}{2\xi} \int_x^{+\infty}
  [a^2(s) - c^2(s)] ds - \frac{1}{2\xi}  \int_0^x (b^2(s) - m^2) ds  -
  \frac{\nu^\pm(\xi)}{\xi}  \int_0^x  c(s) ds + C(\xi),
\end{equation}
where $C(\xi)$ is a constant of integration.
\end{remark}


With this choice, we obtain for $\xi \ne 0$ (see (\ref{Ba2})),
\begin{equation} \label{Rest}
  r^\pm(x,\xi) = (\d_x \phi^\pm)^2 = \frac{1}{4\xi^2}  \big[ a^2(x) + (b^2(x) - m^2) - c^2(x) + d^\pm(x,\xi) \big]^2.
\end{equation}
Moreover it is easy to see that the rests $r^\pm$ satisfy the estimates
\begin{equation} \label{r}
  \forall \alpha, \beta \in \N, \quad |\d_x^\alpha \d_\xi^\beta r^\pm(x,\xi)| \leq
  C_{\alpha \beta} \x^{-2-\alpha} \xsi^{-\beta}, \quad \forall x \in \R^+, \ \forall \xi \in \R^*.
\end{equation}

In our derivation of the phases (\ref{Phase}), it is important to keep in mind that we didn't find an
approximate solution of $B^\pm=0$ but instead of $(B^\pm)^2=0$. Therefore we cannot expect to take $p^\pm = 1$
as a first approximation and we have to work a bit more. So we look for $p^\pm$ such that $B^\pm p^\pm$ be as
small as possible. According to (\ref{Ba1}) and (\ref{Ba2}), we first note that
\begin{equation} \label{Bb0}
  (B^\pm)^2 = r^\pm + 2 (c-\nu^\pm) B^\pm.
\end{equation}
We find now a relation between $B^\pm$ and $(B^\pm)^2$. Using (\ref{B}) and (\ref{Ba4}), we can reexpress
$B^\pm$ as
\begin{equation} \label{Bb1}
  B^\pm = B^\pm_0 + 2 \nu^\pm K^\pm,
\end{equation}
where
\begin{eqnarray}
  B^\pm_0 & = & \Ga \xi + m \Gd - \nu^\pm, \label{B0} \\
  K^\pm & = & \frac{1}{2\nu^\pm} \big[ -\frac{1}{2\xi} (a^2 + (b^2-m^2) - c^2 + d^\pm) \Ga + a\Gb + (b-m) \Gd + c \
  \big].
  \label{K}
\end{eqnarray}
If we take the square of (\ref{Bb1}) we get
\begin{equation} \label{Bb2}
  (B^\pm)^2 = (B^\pm_0)^2 + 2\nu^\pm B^\pm_0 K^\pm + 2\nu^\pm K^\pm B^\pm.
\end{equation}
But from (\ref{B0}) and (\ref{nu}) we see that $(B^\pm_0)^2 = -2\nu^\pm B^\pm_0$. Whence (\ref{Bb2}) becomes
\begin{equation} \label{Bb3}
  (B^\pm)^2 = -2\nu^\pm B^\pm_0 (1-K^\pm) + 2\nu^\pm K^\pm B^\pm.
\end{equation}
Now we replace the expression (\ref{Bb3}) of $(B^\pm)^2$ into (\ref{Bb0}) and we obtain
\begin{equation} \label{Bb4}
  r^\pm = -2\nu^\pm B^\pm_0 (1-K^\pm) + 2\nu^\pm (1+K^\pm-\frac{c}{\nu^\pm}) B^\pm.
\end{equation}
We would like to isolate $B^\pm$ in (\ref{Bb4}). We thus need to invert the functions $(1 + K^\pm -\frac{c}{\nu^\pm})$. Using (\ref{P0H}), (\ref{P0I}) and (\ref{d}), we get the following global asymptotics for $K^\pm$
\begin{equation} \label{EstK0}
  \forall \alpha, \beta \in \N, \quad |\d_x^\alpha \d_\xi^\beta K^\pm(x,\xi)| \leq \left\{\begin{array}{c} C_{\alpha \beta} \x^{-1-\alpha} \xsi^{-1-\beta}, \quad \forall x \in \R^+, \ \forall \xi \in R^*, \\ C_{\alpha \beta} \ \x^{-\alpha} \xsi^{-1-\beta}, \quad \forall x \in \R^-, \ \forall \xi \in R^*. \end{array} \right.
\end{equation}
Let us consider the set $X = \{ \xi \in \R, \ |\xi| \geq R \}$ where $R >> 1$ is a constant. It follows immediately from the asymptotics (\ref{EstK0}) and those of $\frac{c(x)}{\nu^\pm(\xi)}$ that $(1 + K^\pm - \frac{c}{\nu^\pm})$ and $(1-K^\pm)$ are invertible for all $(x,\xi) \in \R \times X$ if the constant $R$ is assumed to be large enough. In consequence we can write (\ref{Bb4}) as
\begin{equation} \label{Bb5}
  B^\pm (1-K^\pm)^{-1} = \frac{1}{2\nu^\pm} (1+K^\pm-\frac{c}{\nu^\pm})^{-1} r^\pm (1-K^\pm)^{-1} + (1+K^\pm-\frac{c}{\nu^\pm})^{-1} B_0^\pm,
\end{equation}
for all $(x,\xi) \in \R \times X$. The first term in the right hand side of (\ref{Bb5}) is small thanks to (\ref{r}) but the second one is not. We choose $p^\pm$ in such a way that they cancel this term. To do this, we observe that the Fourier representations of the projections $P^m_\pm$, \textit{i.e.} the operators
\begin{equation} \label{PNUm}
  P^m_\pm(\xi) = \mathbf{1}_{\R^\pm} \Big( \frac{\xi}{\xi^2 + m^2}(\Ga \xi + m \Gd) \Big) = \frac{1}{2} \Big( I_4 \pm \frac{sgn(\xi)}{\sqrt{\xi^2+m^2}} (\Ga \xi + m\Gd) \Big), \quad \forall \xi \ne 0,
\end{equation}
satisfy the following equations
\begin{equation} \label{Bb6}
  B_0^\pm(\xi) P^m_\pm(\xi) = 0,
\end{equation}
by Lemma \ref{Scalar} and (\ref{B0}). According to (\ref{Bb5}), a natural choice for $p^\pm$ is thus
\begin{equation} \label{p}
  p^\pm = (1-K^\pm)^{-1} P^m_\pm(\xi),
\end{equation}
for which we have
\begin{equation} \label{q}
  q^\pm := B^\pm p^\pm = \frac{1}{2\nu^\pm} (1+K^\pm-\frac{c}{\nu^\pm})^{-1} r^\pm (1-K^\pm)^{-1} P^m_\pm(\xi).
\end{equation}

Let us summarize the situation at this stage. For $\xi \not= 0$, we have defined the phases $\varphi^\pm (x,\xi) = x\xi + \phi^\pm(x,\xi) $ by (\ref{Phase}) and for $\xi \in X$, the amplitudes $p^\pm$ are given by (\ref{p}). Directly from the definitions and from the asymptotics (\ref{P0H}) and (\ref{P0I}) of the potentials $a, b, c$, the following estimates hold.

\begin{lemma}[Estimates on the phases, the amplitudes and related quantities] \label{VariousEst}
For all $ x \in \R^+$ and $\xi \in X$ with $R$ large enough, we have
\begin{equation} \label{EstPhase1}
  \forall \beta \in \N, \quad |\d_\xi^\beta \phi^\pm(x,\xi)| \leq C_{\beta} \log \x \ \xsi^{-\beta}.
\end{equation}
\begin{equation} \label{EstPhase2}
  \forall |\alpha| \geq 1, \forall \beta \in \N,
  \quad |\d_x^\alpha \d_\xi^\beta \phi^\pm(x,\xi)| \leq C_{\alpha \beta} \x^{-\alpha} \xsi^{-\beta}.
\end{equation}
\begin{equation} \label{EstPhase4}
  |\d_{x,\xi}^2 (\varphi^\pm(x,\xi)- x\xi)| \leq \frac{C}{R^2}.
\end{equation}
\begin{equation} \label{EstK}
  \forall \alpha, \beta \in \N, \quad |\d_x^\alpha \d_\xi^\beta K^\pm(x,\xi)|
  \leq C_{\alpha \beta} \x^{-1-\alpha} \xsi^{-1-\beta}.
\end{equation}
\begin{equation} \label{EstP}
  \forall \alpha, \beta \in \N, \quad |\d_x^\alpha \d_\xi^\beta \big(p^\pm(x,\xi) - P^m_\pm(\xi)\big)| \leq
  C_{\alpha \beta} \x^{-1-\alpha} \xsi^{-1-\beta}.
\end{equation}
\begin{equation} \label{EstR}
  \forall \alpha, \beta \in \N, \quad |\d_x^\alpha \d_\xi^\beta r^\pm(x,\xi)
  \leq C_{\alpha \beta} \x^{-2-\alpha} \xsi^{-\beta}.
\end{equation}
\begin{equation} \label{EstQ}
  \forall \alpha, \beta \in \N, \quad |\d_x^\alpha \d_\xi^\beta q^\pm(x,\xi)
  \leq C_{\alpha \beta} \x^{-2-\alpha} \xsi^{-1-\beta}.
\end{equation}
\begin{equation} \label{EstC}
  \forall \alpha, \beta \in \N, \quad |\d_x^\alpha \d_\xi^\beta c^\pm(x,\xi)
  \leq C_{\alpha \beta} \x^{-2-\alpha} \xsi^{-1-\beta}.
\end{equation}
\end{lemma}

Thanks to (\ref{EstPhase1}), (\ref{EstPhase2}), (\ref{EstPhase4})
and (\ref{EstP}), for $R$ large enough, we can define precisely our modifiers $J^{\pm}$ as bounded operators on $\H$
(see \cite{Ro} for instance). Let $\chi^+ \in C^\infty(\R)$ be a cutoff function in space variables
such that $\chi^+(x) = 0$ if $x \leq \frac{1}{2}$ and $\chi^+(x) = 1$ if $x \geq 1$.
Let also $\theta \in C^\infty(\R)$ be a cutoff function in energy variables such that $\theta(\xi) = 0$ if
$|\xi| \leq \frac{1}{2}$ and $\theta(\xi) = 1$ if $|\xi| \geq 1$. For $R$ large enough, $J^{\pm}$ are the Fourier
Integral Operators with phases $\varphi^{\pm} (x, \xi)$ and amplitudes

\begin{equation} \label{Truncatedamplitude}
  P^\pm(x,\xi) = \chi^+ (x) p^\pm(x,\xi)\ \theta(\frac{\xi}{R}).
\end{equation}

We finish this part by a first application of the previous construction. In the next Theorem, the modifiers $J^\pm_{(+\infty)}$ are shown to be time-independent modifications of Isozaki-Kitada type equivalent to the Dollard modification (\ref{Dollard}). Precisely we have
\begin{theorem} \label{Isozaki-Kitada}
 For any $\psi \in \H$ such that $ supp \ \hat{\psi} \subset X$, we have
  \begin{equation} \label{Iso}
    W_{(+\infty)}^\pm \psi = \lim_{t \to \pm \infty} e^{itH} J_{(+\infty)}^\pm
    e^{-it\nu^\pm(D_x)} P^m_\pm \psi.
  \end{equation}
\end{theorem}
\textbf{Proof}: We only sketch the proof for the case $(+)$. By definition of $P^m_+$, we have
\begin{equation} \label{Scalaire}
  U(t) P^m_+ \psi = e^{-it \nu^+ (D_x)}
  e^{ -i \int_0^t [ (b(s { { \mid D_x \mid} \over { \sqrt{D_x^2 +m^2}}} )-m) { m \over {\nu^+(D_x)}} +
  c(s { { \mid D_x \mid} \over { \sqrt{D_x^2 +m^2}}} ) ] ds } P^m_+ \psi := V(t) P^m_+ \psi.
\end{equation}
Then, we write :
\begin{equation}
  e^{itH} J_{(+\infty)}^+ e^{-it\nu^+(D_x)} P^m_+ \psi =  e^{itH} V(t) \big( V^* (t) e^{-it\nu^+(D_x)} \big)
  \big( e^{it\nu^+(D_x)} J_{(+\infty)}^+ e^{-it\nu^+(D_x)} \big) P^m_+ \psi
\end{equation}
\begin{equation} \label{Decomposition}
 \hspace{2cm} = e^{itH} V(t) \  e^{ i \int_0^t [...] ds }\
  \big( e^{it\nu^+(D_x)} J_{(+\infty)}^+ e^{-it\nu^+(D_x)} \big) P^m_+ \psi .
\end{equation}
The classical flow associated with the Hamiltonian $\nu^+ (\xi) = sgn (\xi) \sqrt{ \xi^2 + m^2}$ is given by
\begin{equation} \label{Flot}
  \Phi^t (x, \xi) = (x + t { {\mid \xi \mid} \over {\sqrt{ \xi^2 + m^2}}}, \xi).
\end{equation}
Then, using Egorov's theorem,  we see that $\displaystyle{\big( e^{it\nu^+(D_x)} J_{(+\infty)}^+
e^{-it\nu^+(D_x)} \big)}$ is a FIO with phase $\varphi^+ (t,x,\xi)= x \xi +\phi^+ (x+t\eta, \xi)$, and with
principal symbol\footnote{It means that the others terms of the symbol are $o(1)$ when $t \rightarrow + \infty$.}
$P^+ (x+t\eta, \xi)$ where $\displaystyle{\eta = { { \mid \xi \mid} \over { \sqrt{\xi^2+m^2}}}}$.

\par\noindent
Thus,
$\displaystyle{ e^{ i \int_0^t [...] ds }\ \big( e^{it\nu^+(D_x)} J_{(+\infty)}^+ e^{-it\nu^+(D_x)} \big)}$
is a FIO with the same principal symbol and with phase $\varphi_1^+ (t,x,\xi)= x \xi +\phi_1^+ (t, x, \xi)$ where

$$
\phi_1^+ (t, x, \xi) = \frac{1}{2\xi} \int_{x+t\eta}^{+\infty} [a^2(s) - c^2(s)] ds
                       -\frac{1}{2\xi} \int_0^{x+t\eta} \big[(b^2(s) - m^2) + 2 c(s) \nu^+ (\xi)) \big] ds
$$
\begin{equation} \label{nouvellephase1}
                       + \frac{1}{2\xi} \int_0^{+\infty} (b(s)-m)^2 ds
                       + \int_0^t [(b(s\eta)-m) { m \over {\nu^+ (\xi)}} + c(s\eta) ] ds
\end{equation}
Since $\displaystyle{\frac{1}{2\xi} \int_{x+t\eta}^{+\infty} [a^2(s) - c^2(s)] ds =o(1)}$ when $t \rightarrow + \infty$,
and  by making a change of variables in the last integral, we obtain
$$
\phi_1^+ (t, x, \xi) = -\frac{1}{2\xi} \int_0^{x+t\eta} \big[(b^2(s) - m^2) + 2 c(s) \nu^+ (\xi)) \big] ds
+ \frac{1}{2\xi} \int_0^{+\infty} (b(s)-m)^2 ds
$$
\begin{equation} \label{nouvellephase2}
  + \frac{1}{2\xi} \int_0^{t\eta} [2(b(s)-m) m + 2 c(s) \nu^+ (\xi)  ] ds + o(1) .
\end{equation}
Using again that $\displaystyle{ \int_{t\eta}^{x+t\eta} \big[(b^2(s) - m^2) +
2 c(s) \nu^+ (\xi)) \big] ds = o(1)}$, we see that

$$
\phi_1^+ (t, x, \xi) = -\frac{1}{2\xi} \int_0^{t\eta} \big[(b^2(s) - m^2) + 2 c(s) \nu^+ (\xi)) \big] ds
+ \frac{1}{2\xi} \int_0^{+\infty} (b(s)-m)^2 ds
$$
\begin{equation} \label{nouvellephase3}
  + \frac{1}{2\xi} \int_0^{t\eta} [2(b(s)-m) m + 2 c(s) \nu^+ (\xi)  ] ds + o(1) .
\end{equation}
Then,
\begin{equation} \label{nouvelle phase4}
 \phi_1^+ (t, x, \xi) = - \frac{1}{2\xi} \int_0^{t\eta} (b(s)-m)^2 ds +
 + \frac{1}{2\xi} \int_0^{+\infty} (b(s)-m)^2 ds + o(1) = o(1).
\end{equation}
Using (\ref{EstPhase1}), (\ref{EstPhase2}), (\ref{EstP}) and the continuity of FIOs, we see that
\begin{equation} \label{limite}
  e^{ i \int_0^t [...] ds }\ \big( e^{it\nu^+(D_x)} J_{(+\infty)}^+ e^{-it\nu^+(D_x)} \big) P^m_+ \psi
  = P^m_+ \psi + o(1)
\end{equation}
and Theorem 3.3 follows from (\ref{Decomposition}) and (\ref{limite}). $\diamondsuit$ \\



We now construct the modifiers at high energy $J_{(+\infty)}^\pm(\lambda)$ so that they satisfy
(\ref{Property3}) and (\ref{Prop2}). We still omit the lower index $(+\infty)$ in the next notations.
Comparing (\ref{Prop2}) and (\ref{Iso}) suggests to construct $J^\pm(\lambda)$ close to
$e^{-i\lambda x} J^\pm e^{i\lambda x}$ which are clearly FIOs with phases
$\varphi^\pm(x,\xi,\lambda) = x\xi + \phi^\pm(x,\xi+\lambda)$ and amplitudes $P^\pm(x,\xi + \lambda)$.

With $J^\pm(\lambda) = e^{-i\lambda x} J^\pm e^{i\lambda x}$, we see from (\ref{EstC}) that the amplitudes
$$
  c^\pm (x,\xi,\lambda) = B^\pm(x, \xi + \lambda) P^\pm(x,\xi+ \lambda) -i \Ga  \d_x P^\pm(x,\xi + \lambda),
$$
of the operators $C^\pm(\lambda) = H(\lambda) J^\pm(\lambda) - J^\pm(\lambda) \nu^\pm(D_x+\lambda)$
would satisfy the estimate
\begin{equation} \label{DecayC}
  c^\pm (x,\xi,\lambda) = O(\x^{-2} \lambda^{-1}),
\end{equation}
for $\xi$ in a compact set. Here and in the following, the notation $f(x,\lambda) = O(\x^{-2} \lambda^{-1})$
means that $f(x,\lambda)$ decays as $\x^{-2}$ when $x \to +\infty$ and as $\lambda^{-1}$ when $\lambda \to +\infty$.
We want however the amplitudes $c^\pm(x,\xi,\lambda)$ to be of order $O(\x^{-2} \lambda^{-2})$ and the decay in
(\ref{DecayC}) is not sufficient for our purpose. In consequence, we need to refine our construction.
Following the procedure given in \cite{DN}, we look for modifiers $J^\pm(\lambda)$
defined as FIOs with phases $\varphi^\pm(x,\xi,\lambda)$ and with new amplitudes $P^\pm(x,\xi,\lambda)$
that take the form
\begin{equation} \label{NewAmp}
  P^\pm(x,\xi,\lambda) = \Big[ p^\pm(x,\xi+\lambda) + \frac{1}{\lambda} p^\pm(x,\xi+\lambda) l^\pm(x) +
  \frac{1}{\lambda^2} P_\mp k^\pm(x) \Big] ,
\end{equation}
(up to suitable cutoff functions defined later), where $P_\pm$ denote the projections onto the positive and negative
spectrum of $\Ga$. 
Here the \emph{correctors} $l^\pm, k^\pm$ (that can be matrix-valued) will be functions of $x$ only and should
satisfy some decay in $x$ (see below). It will be clear in the next calculations why we add such correctors to the
amplitudes $p^\pm(x, \xi+\lambda)$.

We now choose $l^\pm$ and $k^\pm$ in (\ref{NewAmp}) so that the amplitudes
\begin{eqnarray} \label{SymbolC}
  c^\pm(x,\xi,\lambda) & = & B^\pm(x,\xi+\lambda) \Big[ p^\pm(x,\xi+\lambda) + \frac{1}{\lambda} p^\pm(x,\xi+\lambda) l^\pm(x) + \frac{1}{\lambda^2} P_\mp k^\pm(x) \Big] \\ &&- i \Ga \Big[ \d_x p^\pm(x,\xi+\lambda) + \frac{1}{\lambda} \d_x p^\pm(x,\xi+\lambda) l^\pm(x) + \frac{1}{\lambda} p^\pm(x,\xi+\lambda) \d_x l^\pm(x) + \frac{1}{\lambda^2} P_\mp \d_x k^\pm(x) \Big], \nonumber
\end{eqnarray}
of the operators $C^\pm(\lambda)$ be of order $O(\x^{-2} \lambda^{-2})$.

To prove this, we need the asymptotics of the different functions appearing in (\ref{SymbolC}).
For $x$ in $\R^+$ and for $\lambda$ large enough, we obtain (after long and tedious calculations)
\begin{equation} \label{AsympNu}
  \nu^\pm(\xi+\lambda) = \pm \big[ \lambda + \xi + \frac{m^2}{2\lambda} \big] + \ O(\lambda^{-2}),
\end{equation}
\begin{equation} \label{AsympD}
  d^\pm(x,\xi+\lambda) = \pm 2 c(x) \big[ \lambda + \xi + \frac{m^2}{2\lambda} \big] + \ O(\x^{-1} \lambda^{-2}).
\end{equation}
\begin{equation} \label{AsympK}
  K^\pm(x,\xi+\lambda) = \pm \frac{1}{2\lambda} \big[ 2 P_{\mp} \, c(x) + a(x) \Gb + (b(x)-m) \Gd \big] + \ O(\x^{-1} \lambda^{-2}).
\end{equation}
\begin{equation} \label{AsympPm}
  P^m_\pm(\xi+\lambda) = P_\pm + \ O(\lambda^{-1}).
\end{equation}
\begin{equation} \label{AsympP}
  p^\pm(x,\xi+\lambda) = P_\pm  + \ O(\lambda^{-1}).
\end{equation}
\begin{equation} \label{AsympDP}
  \d_x p^\pm(x,\xi+\lambda) = \pm \frac{1}{2\lambda} P_\mp (a'(x)\Gb + b'(x)\Gd) + \ O(\x^{-2} \lambda^{-2}).
\end{equation}
\begin{equation} \label{AsympB}
  B^\pm(x,\xi+\lambda) = \mp 2 (\xi+\lambda) P_\mp + 2 c(x) P_\mp + a(x)\Gb + b(x) \Gd + \ O(\lambda^{-1}).
\end{equation}
\begin{equation} \label{AsympQ}
  q^\pm(x,\xi+\lambda) = B^\pm(x,\xi+\lambda) p^\pm(x,\xi+\lambda) =  \pm \frac{1}{2\lambda} c^2(x) P_\pm + \ O(\x^{-2} \lambda^{-2}).
\end{equation}
We mention that the following simple equalities have been used several times to prove the preceding asymptotics
\begin{equation}
  1+\Ga = 2 \left( \begin{array}{cc} I_2&0\\0&0 \end{array} \right) = 2 P_+, \quad 1-\Ga = 2 \left( \begin{array}{cc} 0&0\\0&I_2 \end{array} \right) = 2 P_-.
\end{equation}

By (\ref{AsympP}), (\ref{AsympDP}), (\ref{AsympB}) and (\ref{AsympQ}), the amplitudes $c^\pm(x,\xi,\lambda)$ take the form
\begin{eqnarray*}
  c^\pm(x,\xi,\lambda) & = & \pm \frac{1}{2\lambda} c^2 P_\pm \ \pm \frac{1}{2\lambda^2} c^2 P_\pm l^\pm \\ & & + \frac{1}{\lambda^2} \Big[ \mp 2 (\xi+\lambda) P_\mp + 2 c P_\mp + a\Gb + b \Gd + \ O(\frac{1}{\lambda}) \Big] P_\mp k^\pm  \\ & & - i \Ga \Big[ \pm \frac{1}{2\lambda} P_\mp (a'\Gb + b'\Gd) \pm \frac{1}{2\lambda^2} P_\mp (a'\Gb + b'\Gd) l^\pm \\ & & + \frac{1}{\lambda} \Big( P_\pm + \ O(\frac{1}{\lambda}) \Big) \d_x l^\pm + \frac{1}{\lambda^2} P_\mp \d_x k^\pm \Big]  + \ O(\frac{1}{\x^2 \lambda^2}).
\end{eqnarray*}
From the asymptotics (\ref{P0I}) of the potentials $a,b,c$, we rewrite this last expression as
\begin{equation} \label{SymbolC1}
  c^\pm(x,\xi,\lambda) = \pm \frac{1}{2\lambda} c^2 P_\pm \ \mp \frac{2}{\lambda} P_\mp k^\pm \ \mp \frac{i}{2\lambda} \Ga  P_\mp (a'\Gb + b'\Gd) - \frac{i}{\lambda} \Ga P_\pm \d_x l^\pm + \ R(x,\lambda),
\end{equation}
where the rest $R(x,\lambda)$ satisfies
\begin{equation}\label{Rest1}
  R(x,\lambda) = O\Big( \frac{1 + |l^\pm(x)|}{\x^2 \lambda^2} + \frac{|\d_x l^\pm(x)|}{\lambda^2} +
  \frac{|k^\pm(x)|}{\lambda^2} + \frac{|k^\pm(x)|}{\x \lambda^2} + \frac{|\d_x k^\pm(x)|}{\lambda^2} \Big).
\end{equation}
Now we choose the correctors $l^\pm, k^\pm$ in such a way that the terms of orders
$O(\lambda^{-1})$ in (\ref{SymbolC1}) cancel. Once it is done we shall have to check that
the rest (\ref{Rest1}) be of order $O(\x^{-2} \lambda^{-2})$.

There are clearly two different types of terms in the expression (\ref{SymbolC1}): on one hand the terms
$$
  \pm \frac{1}{2\lambda} c^2 P_\pm - \frac{i}{\lambda} \Ga P_\pm \d_x l^\pm = \frac{1}{\lambda} P_\pm \big[ \pm \frac{1}{2} c^2 \mp i \d_x l^\pm \big],
$$
``live'' in $\H_\pm = P_\pm(\H)$; on the other hand the terms
$$
  \mp \frac{2}{\lambda} P_\mp k^\pm \ \mp \frac{i}{2\lambda} \Ga  P_\mp (a'\Gb + b'\Gd) = \frac{1}{\lambda} P_\mp \big[ \mp 2  k^\pm \ + \frac{i}{2} (a'\Gb + b'\Gd) \big],
$$
``live'' in $\H_\mp = P_\mp(\H)$. Since the Hilbert spaces $\H_-$ and $\H_+$ form a direct sum of $\H$, \textit{i.e.} $\H = \H_- \oplus \H_+$, we can consider separatly the equations
\begin{equation} \label{l}
  \pm \frac{1}{2} c^2 \mp i \d_x l^\pm = 0,
\end{equation}
\begin{equation} \label{k}
  \mp 2 k^\pm + \frac{i}{2} (a'\Gb + b'\Gd) = 0,
\end{equation}
in order to cancel the terms of order $O(\lambda^{-1})$ in (\ref{SymbolC1}). We solve first (\ref{l}) and obtain
\begin{equation} \label{L}
  l^\pm(x) = l(x) = \frac{i}{2} \int_x^{+\infty} c^2(s) ds.
\end{equation}
Then we solve (\ref{k}) and get
\begin{equation} \label{k1}
  k^\pm(x) = \pm \frac{i}{4} (a'(x) \Gb + b'(x) \Gd).
\end{equation}
The functions $l$ and $k^\pm$ clearly satisfy when $x \to +\infty$
\begin{equation}
  l(x) = O(\x^{-1}), \quad \d_x l(x) = O(\x^{-2}), \quad k^\pm(x) = O(\x^{-2}).
\end{equation}
Finally with this choice of correcting terms $l$ and $k^\pm$, we conclude from (\ref{SymbolC1}) and (\ref{Rest1}) that
$$
  c^\pm(x,\xi,\lambda) = R(x,\lambda) = O(\x^{-2} \lambda^{-2}).
$$
In fact, we can prove that for all $x \in \R^+$, $\xi$ in a compact set and $\lambda$ large enough
\begin{equation} \label{EstimatesC}
  \forall \alpha, \beta \in \N, \quad |\d_x^\alpha \d_\xi^\beta c^\pm(x,\xi,\lambda)| \leq C_{\alpha\beta} \  \x^{-2-\alpha} \ \lambda^{-2}.
\end{equation}

Let us summarize the previous results. The modifiers $J^\pm(\lambda)$ are (formally)
constructed as FIOs with phases $\varphi^\pm(x,\xi,\lambda) = x \xi + \phi^\pm(x,\xi+\lambda)$ where
$$
\phi^\pm(x,\xi+\lambda) = \frac{1}{2(\xi+\lambda)} \left( \int_x^{+\infty} [a^2(s) - c^2(s)] ds -
 \int_0^x \big[(b^2(s) - m^2) + d^\pm(s,\xi + \lambda) \ ds \big] \right.
$$
\begin{equation} \label{Phase1}
 \left. + \int_0^{+\infty} (b(s)-m)^2 ds \right),
\end{equation}
and amplitudes
\begin{equation} \label{Amplitude0}
  P^\pm(x,\xi,\lambda) = \Big[ p^\pm(x,\xi+\lambda) + \frac{1}{\lambda} p^\pm(x,\xi+\lambda) l(x) +
  \frac{1}{\lambda^2} P_\mp k^\pm(x) \Big] ,
\end{equation}
where $l$ and $k^\pm$ are given by (\ref{L}) and (\ref{k1}) respectively. \\

\par
Unfortunately, since
$\phi^\pm(x,\xi+\lambda) = O (<x>)$ when $x \rightarrow - \infty$, this phase does not belong to a good class of
oscillating symbols. So, we have to introduce some technical cutoff functions in the amplitude in order to
localize $x$ far away from $-\infty$. Moreover, these cutoff functions must be negligible
in the asymptotics in the previous calculus. We follow the strategy exposed in \cite{Ni2}
which we briefly recall here. \\


We consider a fixed test function $\psi \in \CO$ and we want to calculate  the asymptotics
of $W_{(+\infty)}^\pm(\lambda) \psi$. Since $\hat{\psi} \notin \CO$,
at high energies, translation of wave packets does not dominate over spreading.
So we introduce a cutoff function (depending on $\lambda$) in order to control the spreading. \\

Let
$\chi_0 \in C_0^{\infty } (\R)$ be a cutoff function such that
$\chi_0 (\xi) =1 $ if $\mid \xi \mid \leq 1, \ \chi_0 (\xi) =0 $ if $\mid \xi \mid \geq 2$.
Using the Fourier representation, we have easily :

\begin{equation}\label{energycutoff}
 \forall \epsilon >0, \forall N \geq 1\ ,\
 \mid\mid[ \chi_0 ( {D_x \over {\lambda^{\epsilon}}}) -1] \ \psi \mid \mid_{L^2 (\R )} = O (\lambda^{-N}).
\end{equation}

Now, let us define  the classical propagation zone :

\begin{equation}
\Omega \ =\ \{ x+t \ ;\ x \in supp\ \psi, \ t \in \R^+ \},
\end{equation}
and let $\eta^+ \in C^{\infty}(\R)$ be a cutoff function such that $\eta^+ =1$ in a neighborhood of $\Omega$ and
$\eta^+ =0$ in a neighborhood of $-\infty$. We consider

\begin{equation}\label{Ka}
K^{\pm} (\lambda)\  = (\eta^+ -1) \ e^{-it \nu^{\pm} (D_x +\lambda)} \  \ P_\pm^{m,\lambda} \
\chi_0 ( {D_x \over {\lambda^{\epsilon}}}) \psi
\end{equation}

\begin{lemma} \label{troncakaput}
For $\lambda \gg 1, \ \epsilon \in ]0,1[ , \ t \in \R^{\pm},$ and $N \geq 1 $, we have :
\begin{equation} \label{phnonst}
 \mid \mid K^{\pm} (\lambda)\mid \mid_{L^2 (\R )}  = O(<t>^{-N}  \lambda^{-N} ).
\end{equation}
\end{lemma}
\textbf{Proof}: We only sketch the proof for the case $(+)$. Using the Fourier transform and (\ref{PNUm}), we easily see that
\begin{equation}
K^+(\lambda)\  =  { 1 \over {4\pi}} \ (\eta^+ (x) -1) \ \lambda^{\epsilon} \ \int \left( \int e^{i \varphi
(\xi)} \ \Big( I_4 + \frac{\Ga (\lambda^{\epsilon}\xi + \lambda) + m\Gd}{\sqrt{(\lambda^\epsilon \xi + \lambda)^2 + m^2}} \Big) \ \chi_0 (\xi ) \ d\xi
\right) \psi (y) \ dy ,
\end{equation}
where ${\displaystyle{ \varphi (\xi)= \lambda^{\epsilon} (x-y) \xi - t
\sqrt{ ( \lambda^{\epsilon}\xi + \lambda)^2 +m^2}}}$. So,

\begin{equation}\label{derphase}
\partial_{\xi} \varphi (\xi)=  \lambda^{\epsilon}
\big[ x -(y+  t   { { 1 + \lambda^{\epsilon-1} \xi } \over { \sqrt{ (1 + \lambda^{\epsilon-1} \xi )^2
+ m^2}}}) \big].
\end{equation}
Since $ \xi $ is in a compact set, $\  \epsilon < 1, \ y \in supp \ \psi$,
we easily obtain for $x \in supp \ (\eta^+ -1), and \ \lambda >>1$,

\begin{equation}
\mid  \partial_{\xi} \varphi (\xi) \mid  \geq c \ \lambda^{\epsilon} (1+t),
\end{equation}
for a suitable constant $c>0$. We conclude by a standard non stationary phase argument. $\diamondsuit$  \\

Now, we can define precisely ours modifiers $J^\pm(\lambda)$ in
order to calculate the asymptotics of $W_{(+\infty)}^\pm(\lambda) \psi$. According to (\ref{energycutoff}),
it suffices to calculate the asymptotics  of
${\displaystyle{W_{(+\infty)}^\pm(\lambda) \chi_0 ( {D_x \over {\lambda^{\epsilon}}}) \psi}}$.
We first remark that for $\lambda \gg 1$ and $\epsilon<1$, we have $\xi +\lambda \in X$ if
${\displaystyle{ { \xi \over {\lambda^\epsilon}} \in supp\ \chi_0 }}$.
So, we can define the modifiers $J^\pm(\lambda)$ as FIOs with phases $\varphi^\pm(x,\xi,\lambda)
= x \xi + \phi^\pm(x,\xi+\lambda)$ where $\phi^\pm(x,\xi+\lambda)$ are given by (\ref{Phase1})
and with amplitudes
\begin{equation} \label{Amplitude1}
  P^\pm(x,\xi,\lambda) = \eta^+ (x) \ \Big[ p^\pm(x,\xi+\lambda) + \frac{1}{\lambda} p^\pm(x,\xi+\lambda) l(x) +
  \frac{1}{\lambda^2} P_\mp k^\pm(x) \Big] \
  \chi_0 ( {\xi \over {\lambda^{\epsilon}}}) ,
\end{equation}
where $l$ and $k^\pm$ are given by (\ref{L}) and (\ref{k1}) respectively. \\


 With this definition, we can  mimick the proof of Theorem
 \ref{Isozaki-Kitada},
to get
\begin{lemma}
  For  $\psi \in  \CO$ and for $\lambda$ large, we have
  \begin{equation} \label{ApproximationWO}
    W_{(+\infty)}^\pm(\lambda) \chi_0 ( {D_x \over {\lambda^{\epsilon}}})
    \psi = \lim_{t \to \pm \infty} e^{itH(\lambda)} J_{(+\infty)}^\pm(\lambda) e^{-it\nu^\pm(D_x + \lambda)} P_\pm^{m,\lambda} \psi.
  \end{equation}
\end{lemma}

Moreover, it is easy to see that the estimates (\ref{EstimatesC}) are still satisfied,
so we can prove our main estimate (\ref{Property3}). Precisely we get
\begin{lemma} \label{FundEst}
  For  $\psi \in  \CO$ and when $\lambda$ tends to infinity, the following estimate holds:
  \begin{displaymath}
    \| (W_{(+\infty)}^\pm(\lambda) - J_{(+\infty)}^\pm(\lambda)) \psi \| = O(\lambda^{-2}).
  \end{displaymath}
\end{lemma}
\textbf{Proof}: Everything done in \cite{DN} Lemma 3.3  works here in the same way.
All the contributions coming from the cut-off function $\eta^+$
are negligible using the same arguments as in Lemma \ref{troncakaput} since the
support of the derivatives of $\eta^+$ are far away from $\Omega$. $\diamondsuit$  \\


We end up this section giving the asymptotics of $W^\pm_{(+\infty)}(\lambda)$ when $\lambda$ is large. According to Lemma
\ref{FundEst}, we have for any $\psi \in \COC$, $W^\pm_{(+\infty)}(\lambda) \psi = J^\pm_{(+\infty)}(\lambda) \psi + \ O(\lambda^{-2})$. Thus we only need to compute the asymptotics of the modifier
$J^\pm_{(+\infty)}(\lambda)$ that we shall consider as pseudodifferential operators with symbols
$$
  j^\pm(x,\xi,\lambda) = e^{i\phi^\pm(x,\xi+\lambda)} P^\pm(x,\xi,\lambda).
$$
Using the explicit expressions (\ref{Phase1}) and (\ref{Amplitude1}), we first get the asymptotics
$$
  \phi^\pm(x,\xi+\lambda) = \mp \int_0^x c(s) ds + \frac{1}{2\lambda} \Big[ \int_x^{+\infty} (a^2 - c^2)(s) ds
- \int_0^x (b^2(s) - m^2)  ds
$$
\begin{equation} \label{Phase2}
+ \int_0^{+\infty} (b(s)-m)^2 ds \Big] + \ O(\frac{\log\x}{\lambda^2}),
\end{equation}
\begin{equation} \label{Amplitude2}
  P^\pm(x,\xi,\lambda) = \eta^+ (x) \left[ P_\pm \ \pm \frac{1}{2\lambda} P_\mp \,(a\Gb + b\Gd) + \ \frac{l(x)}{\lambda} P_\pm \right]
+ \ O(\frac{1}{\lambda^2}).
\end{equation}
Moreover using a Taylor expansion of $e^t$ at $t=0$, we get from (\ref{Phase2})
\begin{equation} \label{Phase3}
  e^{i \phi^\pm(x,\xi+\lambda)} = e^{\mp i C^+(x)} \Big[ 1 + \frac{i}{2\lambda} \tilde{C}^+(x) + \ O(\frac{\log\x}{\lambda^2}) \Big],
\end{equation}
with
\begin{equation} \label{C+}
  C^+(x) = \int_0^x c(s) ds,\quad \tilde{C}^+(x) = \int_x^{+\infty} (a^2 - c^2)(s) ds
- \int_0^x (b^2(s) - m^2) ds +  \int_0^{+\infty} (b(s)-m)^2 ds.
\end{equation}
Combining now (\ref{Amplitude2}) and (\ref{Phase3}), we obtain
\begin{equation} \label{Amplitude3}
  j^\pm(x,\xi,\lambda) = e^{\mp i C^+(x)} \ \eta^+ (x)
\Big[ P_\pm \ + \frac{i}{2\lambda} \tilde{C}^+(x) P_\pm \ \pm \frac{1}{2\lambda} P_\mp \,(a\Gb + b\Gd) + \ \frac{l(x)}{\lambda} P_\pm \Big] + \ O(\frac{1}{\lambda^2}).
\end{equation}
But notice from $(\ref{L})$ that
$$
  \frac{i}{2\lambda} \tilde{C}^+(x) + \ \frac{l(x)}{\lambda} = \ \frac{i}{2\lambda}
\Big( \int_x^{+\infty} a^2(s) ds - \int_0^x (b^2(s) - m^2) ds + \int_0^{+\infty} (b(s)-m)^2 ds \Big),
$$
and from the anticommutation properties (\ref{AntiCom}) of the Dirac matrices that
$$
  P_\mp (a\Gb+b\Gd) = (a \Gb + b \Gd) P_\pm.
$$
Hence (\ref{Amplitude3}) becomes
$$
  j^\pm(x,\xi,\lambda) = e^{\mp i C^+(x)} \eta^+ (x) \left[ 1 + \frac{i}{2\lambda}
\Big( \int_x^{+\infty} a^2(s) ds - \int_0^x (b^2(s) - m^2) ds + \int_0^{+\infty} (b(s)-m)^2 ds \right)
$$
\begin{equation} \label{J}
\left. \pm \frac{1}{2\lambda} (a \Gb + b \Gd)  \right] P_{\pm} +O(\frac{1}{\lambda^2}).
\end{equation}
Eventually, if we introduce the notations
\begin{equation}\label{R+}
  R^\pm(x) = \frac{i}{2} \Big( \int_x^{+\infty} a^2(s) ds - \int_0^x (b^2(s) - m^2) ds + \int_0^{+\infty} (b(s)-m)^2 ds
\Big) \ \pm \frac{1}{2}
\,(a\Gb + b\Gd),
\end{equation}
we deduce from (\ref{J}) and the fact that $\eta^+ (x) =1$ on $supp\ \psi$,  the following Proposition
\begin{prop} \label{W+}
    For any $\psi \in \COC$,
  \begin{equation}
    W^\pm_{(+\infty)}(\lambda) \psi = e^{\mp i C^+(x)} \Big[ 1 + \frac{1}{\lambda} R^\pm(x) \Big] P_\pm \psi + \  O(\frac{1}{\lambda^2}),
  \end{equation}
  where $C^+(x)$ and $R^\pm(x)$ are given by (\ref{C+}) and (\ref{R+}) respectively.
\end{prop}


\subsection{Asymptotics of $W_{(-\infty)}^\pm(\lambda)$} \label{AsympWH}

In this subsection, we focus on what happens at the event horizon and give the asymptotics of $W_{(-\infty)}^\pm(\lambda)$ when $\lambda \to +\infty$. In fact, we shall derive them from the results obtained in the preceding subsection \ref{AsympWI} after some simplifications of our model. As usual, we shall omit the lower index $(-\infty)$ in the objects defined or used hereafter.

Recall that the expressions of the wave operators at the event horizon are given by (see (\ref{WOH1}))
$$
  W^\pm = s-\lim_{t \to \pm \infty} e^{itH} e^{-itH_0} P_\mp,
$$
where $H_0 = \Ga D_x + c_0$, $H = \Ga D_x + a \Gb + m \Gd + c$ and the potentials $a, b, c-c_0$ satisfy (\ref{P0H}) when $x \to -\infty$. We first simplify this expression in a convenient way. Let us introduce the unitary transform $U$ on $\H$
\begin{equation} \label{U}
  U = e^{-i\Ga C^-(x)}, \quad C^-(x) = \int_{-\infty}^x [c(s) - c_0] ds + \ c_0 x,
\end{equation}
and define the selfadjoint operators on $\H$
\begin{equation} \label{A0A}
  A_0 = \Ga D_x, \quad A = U^* H U.
\end{equation}
Using (\ref{U}), a short calculation shows that the operator $A$ can be rewritten as
\begin{equation} \label{ExplicitA}
  A = \Ga D_x + W(x),
\end{equation}
where
\begin{equation} \label{PotW}
  W(x) = e^{i\Ga C^-(x)} \big( a(x)\Gb + b(x)\Gd \big) e^{-i\Ga C^-(x)}.
\end{equation}
Note that according to the anticommutation properties (\ref{AntiCom}) of the Dirac matrices, the potential $W$ satisfies $ W\Ga + \Ga W = 0$ and $W^2(x) = a^2(x) + b^2(x)$. Moreover from (\ref{P0H}), we get the following estimates for $W$
\begin{equation} \label{AsympW}
  \exists \alpha >0, \quad W(x) = O(e^{\alpha x}), \quad x \to -\infty.
\end{equation}

Using the unitarity of $U$ and (\ref{A0A}) we rewrite $W^\pm$ as
\begin{eqnarray}
  W^\pm & = & U \, s-\lim_{t \to \pm \infty} e^{itA} U^* e^{-itH_0} P_\mp, \nonumber \\
        & = & U \, s-\lim_{t \to \pm \infty} e^{itA} e^{-itA_0} e^{itA_0} U^* e^{-itH_0} P_\mp. \label{Simpli}
\end{eqnarray}
Now we can simplify the strong limit appearing in (\ref{Simpli}) in two steps. First we claim that
\begin{equation} \label{Int0}
  s-\lim_{t \to \pm \infty} e^{itA_0} U^* e^{-itH_0} P_\mp = e^{i\Ga c_0 x} P_\mp.
\end{equation}
Indeed, using the particular diagonal form of $\Ga$ given in (\ref{DiracMatrices}) and since $e^{-itH_0} = e^{-itA_0} e^{-itc_o}$, we have
\begin{equation}
  e^{itA_0} U^* e^{-itH_0} P_\mp = e^{itA_0} e^{i\Ga C^-(x)} e^{-iA_0} e^{-itc_0} P_\mp =  e^{i\Ga C^-(x \mp t)} e^{-itc_0} P_\mp. \label{Int00}
\end{equation}
When $t \to +\infty$, the right-hand-side of (\ref{Int00}) can be written using (\ref{U}) as
$$
  e^{-i C^-(x - t)} e^{-itc_0} P_- = e^{-i \Big( \int_{-\infty}^{x-t} (c(s) - c_0) ds + c_0 x \Big)} P_-,
$$
from which (\ref{Int0}) follows when $t \to +\infty$. The case $t \to -\infty$ is obtained similarly.

Second since the potential $W$ decays exponentially when $x \to -\infty$ by (\ref{AsympW}), it follows from the methods used in \cite{Da, Me} that the wave operators
\begin{equation} \label{WA0A}
  W^\pm(A,A_0) = s-\lim_{t \to \pm \infty} e^{itA} e^{-itA_0} P_\mp,
\end{equation}
exist on $\H$. Hence by (\ref{Simpli}), (\ref{Int0}), (\ref{WA0A}) and the chain-rule, we obtain the following nice expressions for $W^\pm$
\begin{equation} \label{WHS}
  W^\pm = U \, W^\pm(A,A_0) \, e^{i\Ga c_0 x} \, P_\mp.
\end{equation}
At last since $U$ and $e^{i\Ga c_0 x}$ commute with $e^{i\lambda x}$, it is clear from (\ref{WHS}) that it is enough to know the asymptotics of
$$
  W^\pm(A,A_0,\lambda) = e^{-i\lambda x} W^\pm(A,A_0) e^{i\lambda x}
$$
when $\lambda \to +\infty$ in order to get the asymptotics of $W^\pm(\lambda)$.

Note here that the $\lambda$-shifted wave operator $W^\pm(A,A_0,\lambda)$ is exactly the kind of wave operator studied in our previous paper \cite{DN} in which the asymptotics of $W^\pm(A,A_0,\lambda)$ were calculated. Nevertheless we can also easily derive these asymptotics from the results of the preceding section. For completeness this is what we choose to do here.

We thus follow our usual strategy and construct modifiers $J^\pm_0(\lambda)$ corresponding to $W^\pm(A,A_0,\lambda)$. This problem is in fact similar to the one in subsection \ref{AsympWI}. It suffices to replace $H_0^m$ by $A_0$ and $H$ by $A$ in our calculations. From the explicit form (\ref{A0A}) and (\ref{ExplicitA}) of the operators $A_0$ and $A$, we deduce that we can use the results obtained in subsection \ref{AsympWI} with the following changes: (1) Since the mass $m$ doesn't appear in $A_0$ hence we take $m = 0$. (2) The long-range matrix-valued potential $b$ and scalar potential $c$ don't appear in $A$ (see (\ref{ExplicitA}) and (\ref{AsympW})) hence we put $b(x) = c(x) = 0$. (3) The short-range matrix-valued potential $a(x) \Gb$ is replaced by $W(x)$. (4) The projections $P^m_\pm$ are replaced by $P_\mp$ since we work at the event horizon. Noting that these changes also entail that $\nu^\pm(\xi) = \mp \xi$ and $d^\pm(x,\xi) = 0$, we obtain the following results.

At fixed energy $\lambda=0$, the modifiers $J_0^\pm$ are defined as FIOs with phases
$$
  \varphi^\pm(x,\xi) = x \xi + \frac{1}{2\xi} \int_x^{-\infty} W^2(s) ds,
$$
and amplitudes\footnote{In the same way as the preceding section, we should add some technical cutoff functions which are negligible
in the asymptotics.}
\begin{equation} \label{p-}
  p^\pm(x,\xi) = (1-K^\pm(x,\xi))^{-1} P_\mp, \quad K^\pm(x,\xi) = \mp \frac{1}{2\xi} \Big[ -\frac{W^2(x)}{2\xi} \Ga + W(x) \Big].
\end{equation}
At high energy, the modifiers $J_0^\pm(\lambda)$ are defined as FIOs with phases
\begin{equation} \label{Phase-}
  \varphi^\pm(x,\xi,\lambda) = x \xi + \frac{1}{2(\xi+\lambda)} \int_x^{-\infty} W^2(s) ds,
\end{equation}
and amplitudes
\begin{equation} \label{P-}
  P^\pm(x,\xi,\lambda) = p^\pm(x,\xi+\lambda) + \frac{1}{\lambda^2} P_\pm k^\pm(x),
\end{equation}
where $k^\pm(x) = \mp \frac{i}{4} W'(x)$. Using these definitions and (\ref{AsympW}), we can prove that the symbols $c^\pm(x,\xi,\lambda)$ of the operators $C^\pm(\lambda) = A(\lambda) J^\pm_0(\lambda) - J^\pm_0(\lambda) A_0(\lambda)$ satisfy the estimates
\begin{equation} \label{EstimatesCH}
  \forall \alpha, \beta \in \N, \quad |\d_x^\alpha \d_\xi^\beta c^\pm(x,\xi,\lambda)| \leq C_{\alpha\beta} \  \frac{e^{\alpha x}}{\lambda^2},
\end{equation}
for all $x \in \R^-$ and $\lambda$ large enough. Finally as in the proof of Lemma \ref{FundEst} the estimates (\ref{EstimatesCH}) are the main ingredients to prove the equivalent properties to (\ref{Prop1}) and (\ref{Property3}). Precisely we have
\begin{lemma} \label{FundEst1}
  For any $\psi \in \COC$ and for $\lambda$ large, the following estimate holds
  $$
    \| (W^\pm(A,A_0,\lambda) - J_0^\pm(\lambda)) \psi \| = O(\lambda^{-2}).
  $$
\end{lemma}

We now use Lemma \ref{FundEst1} to compute the asymptotics of $W^\pm(A,A_0,\lambda) \psi$ up to the order $O(\lambda^{-2})$. For any $\psi \in \COC$ and for $\lambda$ large, we have
$$
  W^\pm(A,A_0,\lambda) \psi = J^\pm_0(\lambda) \psi + \ O(\frac{1}{\lambda^2}).
$$
Hence, it is enough to compute the asymptotics of $J_0^\pm(\lambda)$ for $\lambda$ large. Using (\ref{p-}), (\ref{Phase-}), (\ref{P-}) and after some calculations, we obtain
\begin{equation} \label{J0}
  J_0^\pm(\lambda) \psi = \Big[ 1 + \frac{1}{2\lambda} \Big( i \int_x^{-\infty} W^2(s) ds \, \mp \, W(x) \Big) \Big] P_\mp \psi + \ O(\frac{1}{\lambda^2}).
\end{equation}
Note that we retrieve naturally the same formulae as in \cite{DN}. Eventually combining (\ref{WHS}) and (\ref{J0}), we obtain the asymptotics of $W^\pm(\lambda)$ for $\lambda$ large
\begin{prop} \label{W-}
  For any $\psi \in \CO$,
  \begin{equation}
    W^\pm_{(-\infty)}(\lambda) \psi = U \, \Big[ 1 + \frac{1}{\lambda} Q^\pm(x) \Big] e^{i\Ga c_0 x} P_\mp \psi + \ O(\frac{1}{\lambda^2}),
  \end{equation}
  where $U$ is given by (\ref{U}), $Q^\pm(x) =  \frac{1}{2} \Big( i \int_x^{-\infty} W^2(s) ds \, \mp \, W(x) \Big)$ and $W(x)$ is given by (\ref{PotW}).
\end{prop}


\subsection{Proofs of Theorems \ref{MainThm1} and \ref{ReconstructionFormula}}

In this last subsection, we use the asymptotics of $W^\pm_{(\pm \infty)}(\lambda)$ obtained in Propositions \ref{W+} and \ref{W-} to prove the reconstruction formulae given in Theorem \ref{ReconstructionFormula} and finally prove Theorem \ref{MainThm1}. \\

\noindent \textit{Proof of Theorem \ref{ReconstructionFormula}}: We only treat the case of the transmission operator $T_R$ and give the proof of (\ref{RF1}) since the proof of (\ref{RF2}) corresponding to the transmission operator $T_L$ is similar. Recall that we want to compute the asymptotic expansion when $\lambda \to +\infty$ of
$$
  F_l(\lambda) = <T_R e^{i\lambda x} \psi, e^{i\lambda x} \phi> = <W^-_{(+\infty)}(\lambda) \psi,  W^+_{(-\infty)}(\lambda) \phi>,
$$
for $\psi, \phi \in \COC$. Using Propositions \ref{W+} and \ref{W-} and the notations therein, we have
\begin{eqnarray}
  F_l(\lambda) & = & < e^{i C^+(x)} \Big[ 1 + \frac{1}{\lambda} R^-(x) \Big] P_-\psi, U \, \Big[ 1 + \frac{1}{\lambda} Q^+(x) \Big] e^{i \Ga c_0 x} P_- \phi> + \ O(\frac{1}{\lambda^2}), \nonumber \\
                   & = & < e^{i C^+(x)} P_- \psi, U e^{i \Ga c_0 x} P_- \phi> \nonumber \\ & & + \frac{1}{\lambda} \Big[ <e^{i C^+(x)} P_- \psi, U Q^+ e^{i \Ga c_0 x} P_- \phi > + < e^{i C^+(x)} R^- P_- \psi, U e^{i \Ga c_0 x} P_- \phi > \Big] \nonumber \\ & & + \ O(\frac{1}{\lambda^2}). \label{Tl}
\end{eqnarray}
We now compute separatly the terms of different orders in (\ref{Tl}). \\

\noindent\textbf{Order $0$}: Since $\Ga P_- = - P_-$, the term of order $0$ reads
\begin{equation} \label{Order0}
  < e^{i[C^+(x) - C^-(x) + c_0 x]} P_- \psi, P_- \phi >.
\end{equation}
Moreover from (\ref{C+}) and (\ref{U}), the phase $C^+(x) - C^-(x) + c_0$ takes the simple form
\begin{equation} \label{Order0Phase}
  C^+(x) - C^-(x) + c_0 x = -\int_{-\infty}^0 [c(s)-c_0]ds + c_0 x.
\end{equation}

\noindent \textbf{Order $1$}: Using $\Ga P_- = - P_-$ again, the term of order $1$ can be written as
$$
  <e^{i[C^+(x) - C^-(x) + c_0 x]} \, (R^- + (Q^+)^*) \, P_- \psi, P_- \phi >.
$$
Since $W^2 = a^2 + b^2$ and $W P_- = e^{2iC^-} (a \Gb + b \Gd) P_-$ by (\ref{AntiCom}), the term $(Q^+)^* P_-$ takes the form
\begin{equation} \label{Q1}
  (Q^+)^* P_- = \Big( -\frac{i}{2} \int_x^{-\infty} (a^2 + b^2)(s) ds - \frac{1}{2} e^{2iC^-} (a\Gb + b\Gd)  \Big) P_-.
\end{equation}
Moreover from (\ref{R+}) the term $R^-$ is
\begin{equation} \label{R1}
  R^- = \frac{i}{2} \int_x^{+\infty} a^2(s) ds - \frac{i}{2} \int_0^x (b^2(s)-m^2) ds
+ \frac{i}{2} \int_0^{+\infty} (b(s)-m)^2 ds - \half (a\Gb + b\Gd).
\end{equation}
Hence adding (\ref{Q1}) and (\ref{R1}), the term of order $1$ reads
$$
  <e^{i[C^+(x) - C^-(x) + c_0 x]} \Big( \frac{i}{2} \int_{-\infty}^{+\infty} a^2(s) ds +
\frac{i}{2} \int_{-\infty}^0 b^2(s) ds
+ \frac{i}{2} \int_0^{+\infty} (b(s)-m)^2 ds
+ \frac{i}{2} m^2 x \Big) P_- \psi, P_- \phi >
$$
\begin{equation} \label{Q1+R1}
- <e^{i[C^+(x) - C^-(x) + c_0 x]} \Big( \frac{1}{2} e^{2iC^-} (a\Gb + b\Gd) +
\half (a\Gb + b\Gd) \Big) P_- \psi, P_- \phi >.
\end{equation}
Finally using that $e^{i[C^+(x) - C^-(x) + c_0 x]}$ is scalar, that $(a\Gb+b\Gd) P_\pm = P_\mp (a\Gb+b\Gd)$ by (\ref{AntiCom}) and the fact that $<P_+ \psi, P_- \phi> =0$, we see that the last term in (\ref{Q1+R1}) cancel, \textit{i.e.}
$$
  <e^{i[C^+(x) - C^-(x) + c_0 x]} \Big( \frac{1}{2} e^{2iC^-} (a\Gb + b\Gd) +
\half (a\Gb + b\Gd) \Big) P_- \psi, P_- \phi > =0.
$$
Hence the term of order $1$ is
\begin{equation} \label{Order1}
  <e^{i[C^+(x) - C^-(x) + c_0 x]} \Big( \frac{i}{2} \int_{-\infty}^{+\infty} a^2(s) ds +
\frac{i}{2} \int_{-\infty}^0 b^2(s) ds + \frac{i}{2} \int_0^{+\infty} (b(s)-m)^2 ds
+ \frac{i}{2} m^2 x \Big) P_- \psi, P_- \phi >.
\end{equation}
If we introduce the following functions
\begin{eqnarray*}
  \Theta(x) & = & e^{-i\int_{-\infty}^0 [c(s)-c_0]ds + i c_0 x}, \\
  \mathcal{A}(x) & = & \Theta(x) \Big(  \int_{-\infty}^{+\infty} a^2(s) ds +
\int_{-\infty}^0 b^2(s) ds +  \int_0^{+\infty} (b(s)-m)^2 ds +  m^2 x \Big),
\end{eqnarray*}
we have proved the reconstruction formula (\ref{RF1}) and thus Theorem \ref{ReconstructionFormula}. $\diamondsuit$ \\

\noindent \textit{Proof of Theorem \ref{MainThm1}}: We show here that the reconstruction formula (\ref{RF1}) entails the uniqueness of the parameters $M$ and $Q$ under the additional assumption that the charge $q$ of Dirac fields is known, fixed and nonzero. The same result can be shown from the reconstruction formula (\ref{RF2}) in a similar way.

We first compute one of the integrals that appear in (\ref{RF1}) which will be useful in the later analysis. Using the explicit expressions of $F, a_l$ given in (\ref{F}) and (\ref{Potentials}) as well as the definition of the Regge-Wheeler variable $x(r)$ given in (\ref{RWImplicit}), an easy calculation shows that
\begin{equation} \label{IntA}
  \int_\R a_l^2(s) ds = (l+\half)^2 \frac{1}{r_0},
\end{equation}
where $r_0$ is the radius of the event horizon.

Now let us consider two transmission operators $T_{l,1}$ and $T_{l,2}$ corresponding respectively to parameters $M_j, Q_j, m_j, (j=1,2)$ and $q_1 = q_2 = q$ where $q$ is supposed to be known and nonzero. In what follows, all the objects corresponding to $T_{l,j}$ with $j=1,2$ will be denoted by the usual notations with a lower index $j$. We suppose that $T_{l,1} = T_{l,2}$.
In consequence we also have $F_{l,1}(\lambda) = F_{l,2}(\lambda)$. Our goal is to prove that $M_1 = M_2$ and $Q_1 = Q_2$. Using Theorem \ref{ReconstructionFormula} and identifying the terms of same orders in the reconstruction formula (\ref{RF1}), we thus get
\begin{eqnarray}
  \Theta_1(x) & = & \Theta_2(x), \label{ThetaA}\\
  \mathcal{A}_1(x) & = & \mathcal{A}_2(x). \label{A}
\end{eqnarray}
By (\ref{Theta}) and a standard continuity argument, (\ref{ThetaA}) leads to the equality
\begin{equation} \label{Theta1}
  -i\int_{-\infty}^0 [c_1(s)-c_{0,1}]ds + i c_{0,1} x = -i\int_{-\infty}^0 [c_2(s)-c_{0,2}]ds + i c_{0,2} x + \ 2k\pi,
\end{equation}
where $k \in \Z$. If we derivate (\ref{Theta1}) with respect to $x$, we obtain
\begin{equation} \label{C0}
  c_{0,1} = c_{0,2} := c_0.
\end{equation}
Now by (\ref{IntA}), (\ref{A}) leads to the equality
$$
  (l+\half)^2 \frac{1}{r_{0,1}} + \frac{i}{2} \int_{-\infty}^0 b_1^2(s) ds
+ \frac{i}{2} \int_0^{+\infty} (b_1 (s)-m)^2 ds + \frac{i}{2} m_1^2 x
$$
\begin{equation} \label{A1}
= (l+\half)^2 \frac{1}{r_{0,2}} + \frac{i}{2} \int_{-\infty}^0 b_2^2(s) ds
+ \frac{i}{2} \int_0^{+\infty} (b_2 (s)-m)^2 ds + \frac{i}{2} m_2^2 x
\end{equation}
If we derivate (\ref{A1}) with respect to $x$, we first get
\begin{equation} \label{M}
  m_1 = m_2 := m.
\end{equation}
Hence the mass $m$ of Dirac fields is uniquely determined. Moreover, using (\ref{M}), (\ref{IntA}) and the homogeneity in the parameter $l$, we obtain from (\ref{A1})
\begin{equation} \label{R0}
  r_{0,1} = r_{0,2} := r_0.
\end{equation}
Therefore the radius $r_0$ of the event horizon is also uniquely determined. Now if we combine (\ref{R0}) and $c_0 = \frac{qQ}{r_0}$ into (\ref{C0}), we get (since $q$ is supposed to be nonzero)
$$
  Q_1 = Q_2 := Q.
$$
The charge $Q$ of the black hole is thus uniquely determined. Eventually since $r_0$ cancels the function $F$, we get from (\ref{F}) that
$$
  M_1 = M_2 := M = \frac{r_0^2 + Q^2}{2r_0},
$$
and the mass $M$ of the black hole is uniquely determined. This finishes the proof of Theorem \ref{MainThm1}. $\diamondsuit$ \\

\Section{The inverse problem for dS-RN black holes ($\Lambda>0$)} \label{LNot0}

In this Section, we study the inverse problem in the case $\Lambda >0$ corresponding to dS-RN black holes.
In a first part, we prove the same kind of results as in Section \ref{L0}, that is we prove that the parameters
$M, Q$ and $\Lambda$ are uniquely determined by the high energies of the transmission operators $T_L$ or $T_R$.
In a second part, we prove by means of a purely stationary method that the parameters $M, Q$ and $\Lambda$ can also
be uniquely determined from the knowledge of the reflection operators $L$ or $R$ on any interval of energy.


\subsection{The inverse problem at high energy} \label{dSHE}

As in Section \ref{L0}, we shall assume here that one of the following functions of $\lambda \in \R$
$$
  F_l(\lambda) = <T_R e^{i\lambda x} \psi, e^{i\lambda x} \phi>, \quad \quad G_l(\lambda) = <T_L e^{i\lambda x} \psi, e^{i\lambda x} \phi>,
$$
is known for all large values of $\lambda$, for all $l \in \N$ and for all $\psi, \phi \in \H$ with $\hat{\psi},
\hat{\phi} \in \comp(\R; \C^4)$. We emphasize that in this case the construction of the modifiers are simpler
than in the previous section due to the decay of the potentials at infinity; the phases of the modifiers
constructed later will belong to a good class of oscillating symbols. In particular, we do not need a technical
cutoff function $\eta^+$ and a cutoff function $\chi_0$ in order to control the spreading of the wave packets as
in Section \ref{L0} and we can consider test functions $\psi, \phi \in \H$ with $\hat{\psi},
\hat{\phi} \in \comp(\R; \C^4)$. We also assume that the mass $m$ and the charge $q$ of the Dirac fields are known and
fixed. Furthermore the charge $q$ is supposed to be nonzero. Then our main result is
\begin{theorem} \label{MainThm2}
  Under the previous assumptions, the parameters $M, Q$ and $\Lambda$ of the dS-RN black hole are uniquely determined.
\end{theorem}

This Theorem will follow from the following reconstruction formulae obtained on each spin-weighted spherical harmonics
\begin{theorem}[Reconstruction formulae] \label{ReconstructionFormulaNot0}
  Let $\psi, \phi \in \H$ such that $\hat{\psi}, \hat{\phi} \in \comp(\R ; \C^4)$. Then for $\lambda$ large, we have
  \begin{eqnarray}
    F_l(\lambda) & = & <\Theta(x) P_- \psi, P_- \phi> + \frac{1}{\lambda} < \mathcal{A}(x) P_- \psi, P_- \phi> + \ O(\lambda^{-2}), \label{RF1Not0} \\
    G_l(\lambda) & = & <\Theta(x) P_+ \psi, P_+ \phi> - \frac{1}{\lambda} < \mathcal{A}(x) P_+ \psi, P_+ \phi> + \ O(\lambda^{-2}), \label{RF2Not0}
  \end{eqnarray}
  where $\theta(x)$ and $\mathcal{A}(x)$ are multiplication operators given by
  \begin{equation} \label{ThetaANot0}
    \Theta(x) = e^{-i\beta -i(c_+ - c_0)x}, \quad \mathcal{A}(x) = \frac{i}{2} \Big( \int_{-\infty}^{+\infty} \big (a_l^2(s) + b^2(s) \big) ds \Big)\, \Theta(x),
  \end{equation}
  and a constant $\beta$ given by
  $$
    \beta = \int_{-\infty}^0 \big( c(s) - c_0 \big) ds + \int_{0}^{+\infty} \big( c(s) - c_+ \big) ds.
  $$
\end{theorem}
We shall prove Theorem \ref{ReconstructionFormulaNot0} using the same global strategy as in the proof of Theorem
\ref{ReconstructionFormula}. From (\ref{WOH2}), (\ref{WOI2}), (\ref{T}) and the fact that $e^{i\lambda x}$
corresponds to a translation by $\lambda$ in momentum space, we express $F(\lambda)$ and $G(\lambda)$ as follows
\begin{eqnarray}
  F_l(\lambda) & = & <W_{(+\infty)}^-(\lambda) \psi, W_{(-\infty)}^+(\lambda) \phi>,   \label{ExpFNot0} \\
  G_l(\lambda) & = & <W_{(-\infty)}^-(\lambda) \psi, W_{(+\infty)}^+(\lambda) \phi>,  \label{ExpGNot0}
\end{eqnarray}
with
\begin{eqnarray}
  W_{(-\infty)}^\pm(\lambda) & = & e^{-i\lambda x} W_{(-\infty)}^\pm e^{i\lambda x} =  \ s-\lim_{t \to \pm \infty} e^{itH(\lambda)} e^{-itH_0(\lambda)} P_\mp, \label{W-0} \\
  W_{(+\infty)}^\pm(\lambda) & = & e^{-i\lambda x} W_{(+\infty)}^\pm e^{i\lambda x} =  \ s-\lim_{t \to \pm \infty} e^{itH(\lambda)} e^{-it H_+(\lambda)} P_\pm, \label{W++}
\end{eqnarray}
and
$$
  H(\lambda) = \Ga (D_x + \lambda) + a(x) \Gb + b(x) \Gd + c(x),
$$
$$
  H_0(\lambda) = \Ga (D_x + \lambda) + c_0, \quad H_+(\lambda) = \Ga (D_x + \lambda) + c_+.
$$
In consequence, it is enough to obtain an asymptotic expansion of the $\lambda$-shifted wave operators $W_{(\pm\infty)}^\pm(\lambda)$ in order to prove the reconstruction formulae (\ref{RF1Not0}) and (\ref{RF2Not0}).

Note first that the $\lambda$-shifted wave operators $W_{(-\infty)}^\pm(\lambda)$ given by (\ref{W-0}) are exactly the same as in the case $\Lambda = 0$ studied in Subsection \ref{AsympWH}. For completeness we recall here the asymptotic expansion of $W_{(-\infty)}^\pm(\lambda)$ obtained in Proposition \ref{W-}. For any $\psi \in \H$, $\hat{\psi} \in \COC$, we have
\begin{equation} \label{W-Not0}
  W^\pm_{(-\infty)}(\lambda) \psi = U \, \Big[ 1 + \frac{1}{\lambda} Q^\pm(x) \Big] e^{i\Ga c_0 x} P_\mp \psi + \ O(\frac{1}{\lambda^2}),
\end{equation}
where
\begin{equation} \label{U1}
  U = e^{-i\Ga C^-(x)}, \quad C^-(x) = \int_{-\infty}^x [c(s) - c_0] ds + \ c_0 x,
\end{equation}
\begin{equation} \label{U1Bis}
  Q^\pm(x) =  \frac{1}{2} \Big( i \int_x^{-\infty} W^2(s) ds \, \mp \, W(x) \Big), \quad W(x) = e^{i\Ga C^-(x)} \big( a(x)\Gb + b(x)\Gd \big) e^{-i\Ga C^-(x)}.
\end{equation}

Note second that the $\lambda$-shifted wave operators $W_{(+\infty)}^\pm(\lambda)$ given by (\ref{W++}) are very similar to (\ref{W-0}), the constant $c_0$ being replaced by $c_+$ and the projections $P_\mp$ being replaced by $P_\pm$ since we work now at the cosmological horizon. Hence they can be studied exactly the same way as in Section \ref{AsympWH}. Since there are slight modifications in some formulae, we recall here the procedure but omit the proofs. Using the unitary transform (\ref{U1}), we simplify the wave operators $W^\pm_{(+\infty)}$ as follows
\begin{equation}
  W_{(+\infty)}^\pm = U \, s-\lim_{t \to \pm \infty} e^{itA} e^{-itA_0} e^{itA_0} U^* e^{-iH_+} P_\pm,
\end{equation}
where we have used again the notations $A_0 = \Ga D_x$ and $A = U^* H U = \Ga D_x + W(x)$ from (\ref{A0A}) and (\ref{ExplicitA}) with the potential $W$ given by (\ref{U1Bis}). We also recall that by (\ref{AntiCom}) this new potential $W(x)$ satisfies the properties
\begin{equation} \label{WProp}
  \Ga W + W \Ga = 0, \quad W^2 = a^2 + b^2.
\end{equation}
as well as the global estimate
\begin{equation} \label{AsympWNot0}
  \exists \alpha > 0, \quad W(x) = O(e^{-\alpha |x|}), \quad \forall x \in \R.
\end{equation}
The potential $W$ is thus very short-range both at the event horizon and at the cosmological horizon. Now an easy calculation shows that (to be compared with (\ref{Int0}) and its proof)
\begin{equation} \label{Int1}
  s-\lim_{t \to \pm \infty} e^{itA_0} U^* e^{-iH_+} P_\pm = e^{i\Ga \beta} e^{i\Ga c_+ x} P_\pm,
\end{equation}
where the constant $\beta$ is given by
\begin{equation} \label{Beta}
  \beta = \int_{-\infty}^0 \big( c(s) - c_0 \big) ds + \int_{0}^{+\infty} \big( c(s) - c_+ \big) ds.
\end{equation}
Furthermore it is immediate from (\ref{AsympWNot0}) that the wave operators $ W^\pm(A,A_0) = s-\displaystyle\lim_{t \to \pm \infty} e^{itA} e^{-itA_0}$ exist on $\H$. Hence we conclude by the chain-rule that $W_{(+\infty)}^\pm$ take the nice form (to be compared to the expressions (\ref{WHS}) obtained for $W_{(-\infty)}^\pm$)
\begin{equation} \label{WIS}
  W_{(+\infty)}^\pm = U \, W^\pm(A,A_0) \, e^{i\Ga \beta} e^{i\Ga c_+ x} \, P_\pm.
\end{equation}
Since $U$ and $e^{i\Ga \beta} e^{i\Ga c_+ x}$ commute with $e^{i\lambda x}$, we finally get the following expression for $W_{(+\infty)}^\pm(\lambda)$
$$
  W_{(+\infty)}^\pm(\lambda) = U \, W^\pm(A,A_0, \lambda) \, e^{i\Ga \beta} e^{i\Ga c_+ x} \, P_\pm,
$$
where
$$
  W^\pm(A,A_0,\lambda) = e^{-i\lambda x} W^\pm(A,A_0) e^{i\lambda x}.
$$
Clearly it is enough to know the asymptotics of $W^\pm(A,A_0,\lambda) P_\pm$ when $\lambda \to +\infty$ in order to get the asymptotics of $W_{(+\infty)}^\pm(\lambda)$. In fact the calculations are exactly the same to what has been done in subsection \ref{AsympWH} (it suffices to replace $P_\mp$ by $P_\pm$ in these calculations) or in \cite{DN}. Hence we only give the final result without more details. For any $\psi \in \H$, $\hat{\psi} \in \COC$, we finally obtain
\begin{equation} \label{W+Not0}
  W^\pm_{(+\infty)}(\lambda) \psi = U \, \Big[ 1 + \frac{1}{\lambda} \tilde{Q}^\pm(x) \Big] e^{i\Ga \beta} e^{i\Ga c_+ x} P_\pm \psi + \ O(\frac{1}{\lambda^2}),
\end{equation}
where $U$ is given by (\ref{U1}), $\tilde{Q}^\pm(x) =  \frac{1}{2} \Big( i \int_x^{+\infty} W^2(s) ds \, \pm \, W(x) \Big)$ and $W$ is given by (\ref{U1Bis}). \\

\noindent \textit{Proof of Theorem \ref{ReconstructionFormulaNot0}}: We now use the asymptotic expansions (\ref{W-Not0}) and (\ref{W+Not0}) to prove the reconstruction formulae (\ref{RF1Not0}) and (\ref{RF2Not0}). Since the proofs are analogous, we only treat (\ref{RF1Not0}). Using the previous notations we clearly have
\begin{equation} \label{D0}
  F_l(\lambda) = < U \, \Big[ 1 + \frac{1}{\lambda} \tilde{Q}^-(x) \Big] e^{i\Ga \beta} e^{i\Ga c_+ x} P_- \psi, U \, \Big[ 1 + \frac{1}{\lambda} Q^+(x) \Big] e^{i\Ga c_0 x} P_- \phi > + \ O(\frac{1}{\lambda^2}).
\end{equation}
Since $U$ is unitary and since $\Ga P_- = - P_-$, we reexpress (\ref{D0}) as
\begin{eqnarray}
  F(_l\lambda) & = & <e^{-i\beta -i(c_+ - c_0)x} P_- \psi, P_- \phi> \label{D1} \\
      &  & + \frac{1}{\lambda} < e^{-i\beta -i(c_+ - c_0)x} \big( \tilde{Q}^- (x) + (Q^+)^*(x) \big) P_- \psi, P_- \phi> + \ O(\frac{1}{\lambda^2}). \nonumber
\end{eqnarray}
From the explicit expressions of $Q^+$ and $\tilde{Q}^-$, (\ref{D1}) becomes
\begin{eqnarray}
  F_l(\lambda) & = & <e^{-i\beta -i(c_+ - c_0)x} P_- \psi, P_- \phi> \label{D2} \\
      &  & + \frac{1}{\lambda} < e^{-i\beta -i(c_+ - c_0)x} \Big( \frac{i}{2} \int_{-\infty}^{+\infty} W^2(s) ds - W(x) \Big) P_- \psi, P_- \phi> + \ O(\frac{1}{\lambda^2}). \nonumber
\end{eqnarray}
Eventually observe that $W(x) P_- = P_+ W(x)$ by (\ref{AntiCom}) and that $<P_+ \psi, P_- \phi> = 0$. Hence we obtain for (\ref{D2})
\begin{eqnarray}
  F_l(\lambda) & = & <e^{-i\beta -i(c_+ - c_0)x} P_- \psi, P_- \phi> \label{D3} \\
      &  & + \frac{i}{2\lambda} \int_{-\infty}^{+\infty} W^2(s) ds \, < e^{-i\beta -i(c_+ - c_0)x}  P_- \psi, P_- \phi> + \ O(\frac{1}{\lambda^2}). \nonumber
\end{eqnarray}
Denoting
$$
  \Theta(x) = e^{-i\beta -i(c_+ - c_0)x}, \quad \mathcal{A}(x) = \frac{i}{2} \Big( \int_{-\infty}^{+\infty} W^2(s) ds \Big)\, \Theta(x) = \frac{i}{2} \Big( \int_{-\infty}^{+\infty} \big(a_l^2(s) + b^2(s) \big) ds \Big)\, \Theta(x),
$$
we have proved the reconstruction formula (\ref{RF1Not0}). This finishes the proof of Theorem \ref{ReconstructionFormulaNot0}. $\diamondsuit$ \\

\noindent\textit{Proof of Theorem \ref{MainThm2}}: We prove here that the parameters $M, Q$ and $\Lambda$ are uniquely determined from the knowledge of the high energies of the transmission operator $T_R$. Note that the proof with the high energies of $T_L$ is the same. Consider $T_{R,1}$ and $T_{R,2}$ two transmission operators corresponding to parameters $M_j, Q_j, \Lambda_j$ with $j=1,2$ where moreover $m,q \ne 0$ are supposed to be known and fixed. In what follows, we shall denote all the objects associated to $T_{R,j}$ by the usual notations with a lower index $j$.

We assume that $T_{R,1} = T_{R,2}$. From the definition of $F_l(\lambda)$ it follows then that $F_{l,1}(\lambda) = F_{l,2}(\lambda)$. We identify now the terms of same orders in the asymptotic expansion (\ref{RF1Not0}). Since $\psi, \phi$ are dense in $\H$, we get
\begin{eqnarray}
  \Theta_1(x) & = & \Theta_2(x), \quad \forall x \in \R\label{Ord0}\\
  \mathcal{A}_1(x) & = & \mathcal{A}_2(x), \quad \forall x \in \R. \label{Ord1}
\end{eqnarray}
Let us analyse the term of order 0 first. From (\ref{Ord0}) and (\ref{ThetaANot0}), we have
\begin{equation} \label{F0}
  -i\beta_1 -i(c_{+,1} - c_{0,1})x = -i\beta_2 -i(c_{+,2} - c_{0,2})x + 2k\pi, \ \forall x \in \R,
\end{equation}
where $k \in \Z$. If we derivate (\ref{F0}) with respect to $x$, we thus obtain
\begin{equation} \label{F1}
  c_{0,1} - c_{+,1} = c_{0,2} - c_{+,2}.
\end{equation}
Hence using (\ref{F1}) and (\ref{C0C+}), we see that the quantity
\begin{equation} \label{X}
  X = c_0 - c_+ = qQ \frac{r_+ - r_0}{r_0 r_+},
\end{equation}
is uniquely determined.

We analyse now the term of order $O(\lambda^{-1})$. From (\ref{Ord1}), (\ref{ThetaANot0}) and (\ref{Ord0}) again, we have
\begin{equation} \label{F2}
  \int_{-\infty}^{+\infty} W_1^2(s) ds = \int_{-\infty}^{+\infty} W_2^2(s) ds.
\end{equation}
Using that $W^2(x) = a_l^2(x) + b^2(x)$ and the expressions of the potentials $a_l$ and $b$ given by (\ref{Potentials}) and the definition of the Regge-Wheeler variable (\ref{RWImplicit}), we can compute explicitely the integrals that appear in (\ref{F2}). In fact we have
\begin{equation} \label{F3}
  \int_{-\infty}^{+\infty} W^2(s) ds = (l+\half)^2 \Big( \frac{1}{r_0} - \frac{1}{r_+} \Big) + m^2 (r_+ - r_0).
\end{equation}
By homogeneity in $l$ and since $m$ is considered as known and fixed, we deduce from (\ref{F2}) and (\ref{F3}) that
\begin{eqnarray}
  \frac{r_{+,1} - r_{0,1}}{r_{0,1} r_{+,1}} & = & \frac{r_{+,2} - r_{0,2}}{r_{0,2} r_{+,2}}, \\
  r_{+,1} - r_{0,1} & = & r_{+,2} - r_{0,2}.
\end{eqnarray}
Hence the quantities
\begin{equation} \label{YZ}
  Y = \frac{r_+ - r_0}{r_0 r_+}, \quad Z = r_+ - r_0,
\end{equation}
are uniquely determined.

We can now show the uniqueness of the parameters $M, Q$ and $\Lambda$ as follows. We first note the following relation
\begin{equation} \label{F4}
  X = qQ Y.
\end{equation}
Since $X, Y$ are uniquely determined and $q$ is supposed to be known and fixed, we deduce from (\ref{F4}) that $Q$ is uniquely determined, \textit{i.e.} $Q_1 = Q_2 = Q$.

Moreover, from (\ref{YZ}) we deduce that $r_+ - r_0$ and $r_0 r_+$ are uniquely determined. Hence so are $r_0$ and $r_+$ as the unique solutions of the obvious polynomial of second order. Now recall $r_0$ and $r_+$ are roots of $F(r)=0$. The equations $F(r_0) = 0$ and $F(r_+)=0$ can be written using (\ref{F}) as the linear system
\begin{equation} \label{F5}
  \left( \begin{array}{cc} \frac{2}{r_+}& \frac{r_+^2}{3}\\ \frac{2}{r_0}&\frac{r_0^2}{3} \end{array} \right) \ \left( \begin{array}{c} M \\ \Lambda \end{array} \right) = \left( \begin{array}{c} 1+\frac{Q^2}{r_+^2} \\ 1 + \frac{Q^2}{r_0^2} \end{array} \right).
\end{equation}
The determinant of (\ref{F5}) is $\frac{2}{3}\frac{r_0^3 - r_+^3}{r_0 r_+}$ and is clearly nonzero. Hence $(M,\Lambda)$ are the unique solutions of the system (\ref{F5}) whose coefficients depend only on $r_0, r_+, Q$ which are uniquely determined by the previous discussion. We thus conclude that $M$ and $\Lambda$ are also uniquely determined, \textit{i.e.} $M_1 = M_2$ and $\Lambda_1 = \Lambda_2$ and the proof of Theorem \ref{MainThm2} is finished. $\diamondsuit$ \\


\subsection{The inverse problem on an interval of energy} \label{dSAE}

In this last subsection, we solve the inverse problem when the reflection operators $L$ or $R$ are supposed to be known on a (possibly small) interval of energy. We follow the usual stationary approach of inverse scattering on the line and refer to \cite{F, DT} for a presentation of the general method in the case of one-dimensonal Schr\"odinger operators and to \cite{AKM} for an application to massless Dirac operators (see also \cite{G,HJKS} for massive Dirac operators). We first determine a stationary representation of the scattering operator $S$ expressed in terms of the usual transmission and reflection coefficients (here matrices). We do this by a serie of simplications of our model which finally reduces to the exact framework studied in \cite{AKM}. We then use the exponential decay of the potentials to show that the reflection coefficients $R$ and $L$ can be extended analytically to a small strip around the real axis. In consequence, the reflection coefficients $R$ or $L$ are uniquely determined on $\R$ if they are known on any interval of energy by analytic continuation. At last, we use the results of \cite{AKM}, a classical Marchenko method, to prove that the parameters $M, Q$ and $\Lambda$ are uniquely determined by the knowledge of $R(\xi)$ or $L(\xi)$ for all energies.


Recall that the scattering operator $S$ is defined by
$$
  S = (W^+)^* W^-,
$$
where the global wave operators $W^\pm$ are given when $\Lambda > 0$ by
\begin{equation} \label{N0}
  W^\pm = W^\pm_{(-\infty)} + W^\pm_{(+\infty)},
\end{equation}
with
\begin{equation} \label{N1}
  W^\pm_{(-\infty)} = s-\lim_{t \to \pm \infty} e^{itH} e^{-itH_0} P_\mp, \quad \quad W^\pm_{(+\infty)} = s-\lim_{t \to \pm \infty} e^{itH} e^{-itH_+} P_\pm.
\end{equation}
We now use the unitary transform $U$ introduced in (\ref{U}) and the corresponding simplified expressions of $W^\pm_{(\pm \infty)}$ obtained in (\ref{WHS}) and (\ref{WIS}) to express (\ref{N0}) as
\begin{equation} \label{N2}
 W^\pm = U W^\pm(A,A_0) \Big(e^{i\Ga c_0 x} P_\mp + e^{i\Ga \beta} e^{i\Ga c_+ x} P_\pm \Big).
\end{equation}
Here we have used the notations introduced in Subsections \ref{AsympWH} and \ref{dSHE}. Let us denote by $G_\pm$ the operators $e^{i\Ga c_0 x} P_\mp + e^{i\Ga \beta} e^{i\Ga c_+ x} P_\pm$ appearing in (\ref{N2}) and by $S(A,A_0)$ the scattering operator associated to the operators $A$ and $A_0$, \textit{i.e.}
$$
  S(A,A_0) = (W^+(A,A_0))^* W^-(A,A_0).
$$
Using the unitarity of $U$ we thus immediately get the following expression for the scattering operator $S$
\begin{equation} \label{N3}
  S = G_+^* S(A,A_0) G_-.
\end{equation}
The couple of operators $(A,A_0)$ acting on $\H$ turns out to fit the framework studied in \cite{AKM}. Recall
that they are given by $A_0 = \Ga D_x$ and $A = A_0 + W(x)$ where the potential $W(x) = e^{i\Ga C_-(x)} (a(x)
\Gb + b(x) \Gamma^0) e^{-i\Ga C_-(x)}$ is the $4\times$ matrix-valued function
\begin{equation} \label{WA}
  W(x) = \left[ \begin{array}{cc} 0&k(x)\\k^*(x)&0 \end{array} \right],
  \quad k(x) = e^{2iC_-(x)} \left( \begin{array}{cc} -ib(x)&a(x)\\-a(x)&ib(x) \end{array} \right).
\end{equation}
Here $k^*(x)$ denotes the transpose conjugate of the matrix-valued function $k(x)$. Moreover $W$ satisfies (\ref{WProp}) and (\ref{AsympWNot0}) and thus its entries belong to $L^1(\R)$. This is precisely the kind of operators studied in \cite{AKM}. Note however that our potential $W$ is better than $L^1(\R)$ since it is exponentially decreasing at both ends $x \to \pm \infty$. This will be used hereafter. As a consequence, we can use the following stationary representation of $S(A,A_0)$ obtained in \cite{AKM}. Let us introduce the unitary transform $\F$ on $\H$ defined by
\begin{equation}
  \F \psi(\xi) = \frac{1}{\sqrt{2\pi}} \int_\R e^{-i\Ga x\xi} \psi(x) dx,
\end{equation}
then we have (see \cite{AKM}, p 143))
\begin{equation}
  S(A,A_0) = \F^* S_0(\xi) \F,
\end{equation}
where the scattering matrix $S_0(\xi)$ takes the form
\begin{equation}
  S_0(\xi) = \left( \begin{array}{cc} T_L(\xi)&R(\xi)\\L(\xi)&T_R(\xi) \end{array} \right).
\end{equation}
Here $T_L(\xi)$ and $T_R(\xi)$ are $2\times2$ matrix-valued functions which correspond to the usual transmission coefficients of $S$ whereas $L(\xi)$ and $R(\xi)$ are $2\times2$ matrix-valued functions which correspond to the usual reflection coefficients of $S$. We refer to Sections 2 and 3 of \cite{AKM} for the definition and the construction of the scattering matrix $S_0(\xi)$. Hence (\ref{N3}) becomes
\begin{equation} \label{N4}
  S = (\F G_+)^* S_0(\xi) \F G_-.
\end{equation}
We now finish our factorization of the scattering operator $S$ as follows. Using $2\times2$ block matrix notations, we note that
$$
  G_+ = \left( \begin{array}{cc} e^{i\beta}&0\\0&1 \end{array} \right) \left( \begin{array}{cc} e^{ic_+ x}&0\\0&e^{-ic_0 x} \end{array} \right), \quad \quad G_- = \left( \begin{array}{cc} 1&0\\0&e^{-i\beta} \end{array} \right) \left( \begin{array}{cc} e^{ic_0 x}&0\\0&e^{-ic_+ x} \end{array} \right),
$$
and we define two unitary transforms $F_\pm$ on $\H$ by
\begin{equation} \label{F+}
  F_+ \psi(\xi) = \F \left( \begin{array}{cc} e^{ic_+ x}&0\\0&e^{-ic_0 x} \end{array} \right) \psi(\xi) = \frac{1}{\sqrt{2\pi}} \int_\R  \left( \begin{array}{cc} e^{-ix\xi + ic_+ x}&0\\0&e^{ix\xi - ic_0 x} \end{array} \right) \psi(x) dx,
\end{equation}
and
\begin{equation} \label{F-}
  F_- \psi(\xi) = \F \left( \begin{array}{cc} e^{ic_0 x}&0\\0&e^{-ic_+ x} \end{array} \right) \psi(\xi) = \frac{1}{\sqrt{2\pi}} \int_\R  \left( \begin{array}{cc} e^{-ix\xi + ic_0 x}&0\\0&e^{ix\xi - ic_+ x} \end{array} \right) \psi(x) dx.
\end{equation}
Then we have
\begin{equation} \label{N5}
  \F G_+ = \left( \begin{array}{cc} e^{i\beta}&0\\0&1 \end{array} \right) F_+, \quad \quad \F G_- = \left( \begin{array}{cc} 1&0\\0&e^{-i\beta} \end{array} \right) F_-.
\end{equation}
Hence we conclude from (\ref{N5}) that the scattering operator (\ref{N4}) factorizes as
\begin{equation} \label{N6}
  S = F_+^* \left( \begin{array}{cc} e^{-i\beta} T_L(\xi)&e^{-2i\beta} R(\xi)\\ L(\xi)&e^{-i\beta} T_R(\xi) \end{array} \right) F_-.
\end{equation}
We summarize this result as a Proposition
\begin{prop} \label{ScatMat}
  The scattering operator $S$ has the following stationary representation. If $F_\pm$ are the unitary transforms defined in (\ref{F+}) and (\ref{F-}), then
  \begin{equation} \label{StatRep}
    S = F_+^* S(\xi) F_-,
  \end{equation}
  where the $4\times4$ scattering matrix $S(\xi)$ is given by
  \begin{equation} \label{ScatMatrix}
    S(\xi) = \left( \begin{array}{cc} e^{-i\beta} T_L(\xi)&e^{-2i\beta} R(\xi)\\ L(\xi)&e^{-i\beta} T_R(\xi) \end{array} \right),
  \end{equation}
  and the quantities $T_L, T_R$ and $L, R$ are the $2\times2$ matrices that correspond to the transmission and reflection matrices of $S(A,A_0)$ respectively and are obtained in \cite{AKM}, Sections 2 and 3.
\end{prop}
\begin{remark}
  As the notations suggest, the diagonal elements of the scattering matrix $S(\xi)$ given in (\ref{ScatMatrix}) are simply the stationary representations of the transmission operators $T_L$ and $T_R$ introduced in Section \ref{dS-RN}, (\ref{T}). The anti-diagonal elements of $S(\xi)$ are in turn the stationary representations of the reflection operators $L$ and $R$ in (\ref{R}).
\end{remark}
\begin{remark}
  The unitary operators $F_\pm$ appearing in the stationary representation (\ref{StatRep}) of $S$ are natural in the following sense. Let us define the two selfadjoint operators on $\H$
  $$
    H^+ = (\Ga D_x + c_+) P_+ + (\Ga D_x + c_0) P_-, \quad \quad H^- = (\Ga D_x + c_0) P_+ + (\Ga D_x + c_+) P_-.
  $$
  Hence it is clear from (\ref{N0}) and (\ref{N1}) that the global wave operators can be written in a classical form as
  $$
    W^\pm = s-\lim_{t \to \pm \infty} e^{itH} e^{-itH^\pm}.
  $$
  Now it is an easy calculation to show that the unitary transforms $F_\pm$ introduced in (\ref{F+}) and (\ref{F-}) are precisely the unitary transforms which diagonalize the operators $H^\pm$, \textit{i.e}
  $$
    H^\pm = F_\pm^* M_\xi F_\pm,
  $$
where $M_\xi$ denotes the multiplication operator by $\xi$. We conclude that (\ref{StatRep}) together with (\ref{ScatMatrix}) are the expected stationary representation of the scattering operator $S$.
\end{remark}


In the sequel, we shall use the explicit link between our scattering matrix $S(\xi)$ and the scattering matrix $S_0(\xi)$ thoroughly studied in \cite{AKM} in order to solve the inverse problem. Let us first briefly summarize some of the main results obtained in \cite{AKM}. Under the assumption $W \in L^1(\R)$, the scattering matrix $S_0(\xi)$ is continuous for $\xi \in \R$ and tends to $I_4$ when $\xi \to \pm \infty$. It is also unitary for each $\xi \in \R$ (see \cite{AKM}, Thm 3.1 for a proof of these statements and for other properties on $S_0(\xi)$)). Moreover, the following partial characterization result holds:
\begin{theorem} [\cite{AKM}, Thm 6.3] \label{Reconstruction}
  Assume that the reflection operators $R(\xi)$ and $L(\xi)$ be $2\times2$ matrix valued functions satisfying
  \begin{equation} \label{Cond1}
    \sup_{\xi \in \R} \| R(\xi) \| < 1, \quad \sup_{\xi \in \R} \| L(\xi) \| < 1, \quad \quad \| \hat{R}(\alpha)\| \in L^1(\R), \quad \| \hat{L}(\alpha) \| \in L^1(\R),
  \end{equation}
  \begin{equation} \label{Cond2}
    \int_0^{+\infty} \alpha \| \hat{R}(\alpha) \|^2 d\alpha < \infty, \quad \int_{-\infty}^0 \alpha \| \hat{L}(\alpha) \|^2         d\alpha < \infty,
  \end{equation}
  where $\hat{R}(\alpha)$ and $\hat{L}(\alpha)$ denote the usual Fourier transform of $R(\xi)$ and $L(\xi)$ and $\|.\|$ is the euclidean norm of a given matrix. Then the matrix-valued function $k(x) \in L^1(\R)$ in (\ref{WA}) (and thus the potential $W(x)$) can be uniquely recovered from the knowledge of $R(\xi)$ and $L(\xi)$ for all $\xi \in \R$.
\end{theorem}
We make several comments on this result and how we can apply it to our model:
\begin{itemize}
\item The proof of the above theorem uses a classical Marchenko method. For instance, the matrix-valued function $k(x)$ can be obtained after solving the following Marchenko integral equations for $\alpha >0$ (see \cite{AKM}, eq. (6.9) and (6.11))
\begin{equation} \label{Mar1}
  B_1(x,\alpha) = -\hat{R}(\alpha+2x) + \int_0^{+\infty} \int_0^{+\infty} B_1(x,\gamma) \hat{R}(\delta+\gamma+2x)^* \hat{R}(\alpha+ \delta+2x) d\gamma d\delta,
\end{equation}
\begin{equation} \label{Mar2}
  B_2(x,\alpha) = -\hat{L}(\alpha-2x)^* + \int_0^{+\infty} \int_0^{+\infty} B_2(x,\gamma) \hat{L}(\delta+\gamma-2x) \hat{L}(\alpha+ \delta-2x)^* d\gamma d\delta.
\end{equation}
Under the assumption (\ref{Cond1}), the integral equations (\ref{Mar1}) and (\ref{Mar2}) are uniquely solvable in $L^1(\R^+)$ (\cite{AKM}, Thm 6.2). Moreover, under the additionnal assumption (\ref{Cond2}), the matrix-valued function $k(x)$ defined using the boundary values of $B_1$ and $B_2$ by the formulae (see \cite{AKM}, eq. (4.19))
$$k(x) = 2i B_1(x,0^+), \ \forall x>0, \quad k(x) = -2i B_2(x,0^+), \ \forall x<0,
$$
can be shown to be in $L^1(\R)$ and thus corresponds to the potential we are looking for. \\
\item If the potential $W$ belongs to $L^1(\R)$, then the condition (\ref{Cond1}) is automatically satisfied (see \cite{AKM}, Thm 4.2 and eq. (6.17)). Although this condition is the natural one under which one could expect to reconstruct the potential $k$ in the class $L^1$, the authors of \cite{AKM} had to add the extra assumption (\ref{Cond2}) (which must then be checked) in order to prove their result. We refer to \cite{AKM}, p. 154 for more details on this point. In our case, we shall prove the condition (\ref{Cond2}) as follows. Using the exponential decay of $W$, we are first able to show that the reflection coefficients $R(\xi)$ and $L(\xi)$ (in fact the whole scattering matrix $S_0(\xi)$) are analytic on a small strip around the real axis. Moreover the functions $R(.+i\eta)$ and $L(.+i\eta)$ can be shown to belong to $L^2(\R)$ uniformly for each $|\eta|$ small enough. It follows then from standard results on the Fourier transform (see for instance \cite{RS}, Thm IX.13) that $\hat{R}(\alpha)$ and $\hat{L}(\alpha)$ satisfy
$$
  e^{\epsilon |\alpha|} \hat{R}(\alpha) \in L^2(\R), \quad e^{\epsilon |\alpha|} \hat{L}(\alpha) \in L^2(\R), \quad \forall \epsilon \ \textrm{small enough},
$$
from which (\ref{Cond2}) follows immediately. \\
\item From (\ref{Mar1}) and (\ref{Mar2}) and the reconstruction procedure explained above, we see that the knowledge of $R(\xi)$ and $L(\xi)$ for all $\xi \in \R$ is used to recover the potential $k(x)$ for all $x \in \R$. In fact it is only enough to know either $R(\xi)$ or $L(\xi)$ for all $\xi \in \R$ since then the whole scattering matrix $S_0(\xi)$ can be uniquely recovered. The procedure is explained in \cite{AKM}, p.147, eq.(5.3)-(5.5) and we reproduce it for completeness. Assume for instance that $R(\xi)$ is known for all $\xi \in \R$. Then the transmission coefficients $T_L(\xi)$ and $T_R(\xi)$ can be obtained performing the factorizations
\begin{equation} \label{Fact1}
  T_L(\xi) T_L(\xi)^* = I_4 - R(\xi) R(\xi)^*, \quad T_R(\xi)^* T_R(\xi) = I_4 - R(\xi)^* R(\xi), \quad \xi \in \R.
\end{equation}
Under the assumption $k \in L^1(\R)$, it was shown in \cite{AKM} that the above factorization problems are in fact left or right canonical Wiener-Hopf factorization in the Wiener algebra $\mathcal{W}^4$ and thus lead to unique $T_L(\xi)$ and $T_R(\xi)$ (see for instance \cite{GGK}, Thm 9.2, p.831). At last, the reflection coefficient $L(\xi)$ is recovered from $R(\xi)$ by the formula
\begin{equation} \label{Fact2}
  L(\xi) = - T_R(\xi) R(\xi)^*(T_L(\xi)^*)^{-1}.
\end{equation}
\item Eventually we explain how we can apply this result to our model. From Proposition \ref{ScatMat}, we assume
for instance that $e^{-2i\beta} R(\xi)$ is known for all $\xi \in \R$. Then it is easy to see from (\ref{Fact1})
and (\ref{Fact2}) that we can uniquely recover $T_L(\xi)$ and $T_R(\xi)$ by performing Wiener-Hopf
factorizations and then $e^{2i\beta} L(\xi)$ for all $\xi \in \R$. Note that the exponential term $e^{-2i\beta}$
disappears in the factorization (\ref{Fact1}). If we assume that the assumptions (\ref{Cond1}) and (\ref{Cond2})
hold (this will be checked below), then we can apply Thm \ref{Reconstruction} as follows. Multiplying the
integral equations (\ref{Mar1}) and (\ref{Mar2}) by $e^{-2i\beta}$ and solving them, we conclude that we can
uniquely recover $e^{2i\beta} k(x)$ (and not $k(x)$) for all $x\in \R$. We shall show below that this implies
the uniqueness of the parameters $M, Q$ and $\Lambda$ of the black hole.
\end{itemize}


Let us now show the analyticity of $R(\xi)$ and $L(\xi)$ on a small strip around the real axis and prove there the uniform $L^2$ estimates mentioned above. To do this we need to introduce some objects whose existence has been shown in \cite{AKM}, Sections 1, 2 and 3. The reflection coefficients $R(\xi)$ and $L(\xi)$ can be expressed in terms of solutions of the stationary problem
\begin{equation} \label{Stationary}
  \Big[ \Ga D_x + W(x)\Big] X(x,\xi) = \xi X(x,\xi), \quad \xi \in \R
\end{equation}
where $X(x,\xi)$ is understood as $4\times4$ matrix-valued function. Of special interest are the Jost solutions $F_l(x,\xi)$ and $F_r(x,\xi)$ of (\ref{Stationary}) which are singled out by the specific asymptotics at infinity
$$
  F_l(x,\xi) = e^{i\Ga\xi x} (I_4 + o(1)), \quad x \to +\infty,
$$
$$
  F_r(x,\xi) = e^{i\Ga\xi x} (I_4 + o(1)), \quad x \to -\infty.
$$
For each $\xi \in \R$, these two solutions exist, are fundamental matrices of (\ref{Stationary}) and are related as follows (\cite{AKM}, Proposition 2.2). There exist two $4\times4$ matrix valued functions $a_l(\xi)$ and $a_r(\xi)$ such that
$$
  F_l(x,\xi) = F_r(x,\xi) a_l(\xi), \quad F_r(x,\xi) = F_l(x,\xi) a_r(\xi),
$$
and satisfying $a_l(\xi) a_r(\xi) = a_r(\xi) a_l(\xi) = I_4$ for all $\xi \in \R$. Note that $F_l(x,\xi)$ and $F_r(x,\xi)$ satisfy the asymptotics (in the opposite ends)
\begin{equation} \label{Jost}
  F_l(x,\xi) = e^{i\Ga\xi x} (a_l(\xi) + o(1)), \quad x \to -\infty,
\end{equation}
$$
  F_r(x,\xi) = e^{i\Ga\xi x} (a_r(\xi) + o(1)), \quad x \to +\infty.
$$
Let us now express $a_l(\xi)$ and $a_r(\xi)$ using $2\times2$ block matrix notations as
$$
  a_l(\xi = \left[ \begin{array}{cc} a_{l1}(\xi)&a_{l2}(\xi)\\a_{l3}(\xi)&a_{l4}(\xi) \end{array} \right], \quad a_r(\xi = \left[ \begin{array}{cc} a_{r1}(\xi)&a_{r2}(\xi)\\a_{r3}(\xi)&a_{r4}(\xi) \end{array} \right],
$$
Then the reflection coefficients are defined by (\cite{AKM}, eq. (3.6) and (3.7))
$$
  R(\xi) = a_{r2}(\xi) a_{r4}(\xi)^{-1} = - a_{l1}(\xi)^{-1} a_{l2}(\xi),
$$
$$
  L(\xi) = a_{l3}(\xi) a_{l1}(\xi)^{-1} = - a_{r4}(\xi)^{-1} a_{r3}(\xi).
$$
Since the situations are obviously symmetric, we shall only prove the analyticity and the uniform $L^2$ estimate on a small strip around the real axis for $R(\xi)$ (the proof for $L(\xi)$ being identical). Moreover, we shall only consider the definition $R(\xi) = - a_{l1}(\xi)^{-1} a_{l2}(\xi)$ for simplicity. To go further, we use some integral representations of the coefficients $a_{l1}(\xi)$ and $a_{l2}(\xi)$ obtained in \cite{AKM}. These are given in terms of the Faddeev matrix $M_l(x,\xi)$ defined by
$$
  M_l(x,\xi) = F_l(x,\xi) e^{-\Ga\xi x}.
$$
It is easy to see from (\ref{Stationary}) that $M_l(x,\xi)$ must satisfy the integral equation (\cite{AKM}, eq. (2.12))
\begin{equation} \label{FaddeevIntEq}
  M_l(x,\xi) = I_4 - i\Ga \int_x^{+\infty} e^{-i\Ga \xi(y-x)} W(y) M_l(y,\xi) e^{i\Ga \xi (y-x)} dy,
\end{equation}
and from (\ref{Jost}) that $M_l(x,\xi)$ must satisfy the asymptotics $M_l(x,\xi) = I_4 + o(1)$ when $x \to +\infty$. In fact, using once again $2\times2$ block matrix notations for $M_l(x,\xi)$
$$
  M_l(x,\xi) = \left[ \begin{array}{cc} M_{l1}(x,\xi)&M_{l2}(x,\xi)\\M_{l3}(x,\xi)&M_{l4}(x,\xi) \end{array} \right],
$$
and iterating (\ref{FaddeevIntEq}) once, we get the uncoupled system of integral equations for $M_{l3}(x,\xi)$ and $M_{l4}(x,\xi)$ (\cite{AKM}, eq. (2.15) and (2.16))
\begin{equation} \label{IE1}
  M_{l3}(x,\xi) = i \int_x^{+\infty} e^{2i\xi(y-x)} k(y) dy + \int_x^{+\infty} \int_y^{+\infty} e^{2i\xi(y-x)} k(y)^* k(z) M_{l3}(z,\xi) dz dy,
\end{equation}
\begin{equation} \label{IE2}
  M_{l4}(x,\xi) = I_4 + \int_x^{+\infty} \int_y^{+\infty} e^{-2i\xi(z-y)} k(y)^* k(z) M_{l4}(z,\xi) dz dy,
\end{equation}
and similar equations for $M_{l1}(x,\xi)$ and $M_{l2}(x,\xi)$ that we won't need. Eventually, the following integral representations for the coefficients $a_{l1}(\xi)$ and $a_{l2}(\xi)$ hold (\cite{AKM}, eq. (2.25) and (2.26))
\begin{equation} \label{IR1}
  a_{l1}(\xi) = I_2 - i\int_\R k(y) M_{l3}(y,\xi) dy,
\end{equation}
\begin{equation} \label{IR2}
  a_{l2}(\xi) = - i\int_\R e^{-2i\xi y} k(y)^* M_{l4}(y,\xi) dy.
\end{equation}
We first study the coefficient $a_{l2}(\xi)$ expressed in terms of the Faddeev matrix $M_{l4}(x,\xi)$. Under the assumption $k \in L^1(\R)$, a solution $M_{l4}(x,\xi)$ of (\ref{IE2}) with the right asymptotics is easily shown to exist by iteration. Moreover for each fixed $x \in \R$, this solution can be extended to a continuous function in the variable $\xi$ when Im$\xi \leq 0$ and analytic when Im$\xi <0$ (\cite{AKM}, Proposition 2.3). We prove now the following result
\begin{lemma} \label{Ml4}
  Define the function $P(x,\xi) = \int_x^{+\infty} e^{2|Im\xi||y|} \|k(y)\| dy$. Then there exists $\kappa>0$ small enough such that\\
  (i) For all $\xi$ satisfying $|Im\xi| \leq \kappa$ and for all $x \in \R$, the function $P(x,\xi)$ is uniformly bounded.\\
  (ii) For each fixed $x \in \R$, the Faddeev matrix $M_{l4}(x,\xi)$ can be extended analytically to the strip $|Im\xi|<\kappa$. Moreover, for each such $\xi$, it satisfies the estimate
  \begin{equation}
    \|M_{l4}(x,\xi)\| \leq C \cosh(P(x,\xi)).
  \end{equation}
  (iii) For each fixed $x \in \R$, the derivative $M'_{l4}(x,\xi)$ of the Faddeev matrix w.r.t. the variable $x$ can be extended analytically to the strip $|Im\xi|<\kappa$. Moreover, for each such $\xi$, it satisfies the estimate
  \begin{equation}
    \|M'_{l4}(x,\xi)\| \leq C \sinh(P(x,\xi)).
  \end{equation}
\end{lemma}
\textit{Proof}: The first assertion is a direct consequence of the definition of $P(x,\xi)$ and (\ref{AsympWNot0}) (take for instance $\kappa = \frac{\alpha}{2}$ where $\alpha$ is the positive number that appears in (\ref{AsympWNot0})). Solving (\ref{IE2}) by iteration leads to set $M_{l4}(x,\xi) = \sum_{n=0}^\infty u_n(x,\xi)$ with $u_0(x,\xi) = I_2$ and
\begin{equation} \label{Ite1}
  u_n(x,\xi) = \int_x^{+\infty} \int_y^{+\infty} e^{-2i\xi(z-y)} k(y)^* k(z) u_{n-1}(z,\xi) dz dy, \quad \forall n \geq 1.
\end{equation}
By induction we get the estimates
\begin{equation} \label{Es1}
  \|u_n(x,\xi)\| \leq \frac{P(x,\xi)^{2n}}{(2n)!}, \quad \forall n \in \N.
\end{equation}
Together with (i), this entails the second assertion. To prove the third one, we consider the serie of derivatives $\sum_{n=1}^\infty u'_n(x,\xi)$. From (\ref{Ite1}), note that
$$
  u'_n(x,\xi) = - \int_x^{+\infty} e^{-2i\xi(z-x)} k(x)^* k(z) u_{n-1}(z,\xi) dz dy.
$$
By induction and using (\ref{Es1}), we get the estimates $\|u'_n(x,\xi)\| \leq C \frac{P(x,\xi)^{2n-1}}{(2n-1)!}$ for all $n \geq 1$ from which we deduce (iii). \\ $\diamondsuit$ \\
\begin{coro} \label{al2}
  Let $\kappa$ the positive number defined in Lemma \ref{Ml4}. The coefficient $a_{l2}(\xi)$ is analytic on the strip $|Im\xi|<\kappa$. Moreover, it  satisfies there the estimate
\begin{equation}
  \|a_{l2}(\xi)\| = O(|\xi|^{-1}), \quad |\xi| \to \infty.
\end{equation}
\end{coro}
\textit{Proof}: The analyticity on the strip $|Im\xi|<\kappa$ follows directly from (\ref{IR2}) and Lemma \ref{Ml4}. To prove the second assertion, we integrate by parts in (\ref{IR2}). For all $\xi$ with $|Im\xi|<\kappa$, we obtain
\begin{equation} \label{IR3}
  a_{l2}(\xi) = -\frac{1}{2\xi} \int_\R e^{-2i\xi y} \Big( k'(y) M_{l4}(y,\xi) + k(y) M'_{l4}(y,\xi) \Big) dy.
\end{equation}
Since $k'$ also satisfies the estimate (\ref{AsympWNot0}) and using Lemma \ref{Ml4} again, we conclude that $\|a_{l2}(\xi)\| \leq \frac{C}{|\xi|}$. \\ $\diamondsuit$ \\
We now study the coefficient $a_{l1}(\xi)$ expressed in terms of the Faddeev matrix $M_{l3}(x,\xi)$. Once again under the assumption $k \in L^1(\R)$, a solution $M_{l3}(x,\xi)$ of (\ref{IE1}) with the right asymptotics is easily shown to exist by iteration. Moreover for each fixed $x \in \R$, this solution can be extended to a continuous function in the variable $\xi$ when Im$\xi \geq 0$ and analytic when Im$\xi >0$ (\cite{AKM}, Proposition 2.3). Using the same function $P(x,\xi)$ and positive number $\kappa$ as in Lemma \ref{Ml4}, let us prove the following result
\begin{lemma} \label{Ml3}
  For each fixed $x \in \R$, the Faddeev matrix $M_{l3}(x,\xi)$ can be extended analytically to the strip $|Im\xi|<\kappa$. Moreover, for each such $\xi$, it satisfies the estimates
  \begin{equation} \label{Ml3-1}
    \|M_{l3}(x,\xi)\| \leq C e^{2|Im\xi||x|} \sinh(P(x,\xi)).
  \end{equation}
  \begin{equation}\label{Ml3-2}
    \|M_{l3}(x,\xi)\| \leq \frac{C}{|\xi|} \Big(1+ e^{2|Im\xi||x|} \Big), \quad |\xi| \geq 1.
  \end{equation}
\end{lemma}
\textit{Proof}: We solve (\ref{IE1}) by iteration. Hence we set $M_{l3}(x,\xi) = \sum_{n=0}^\infty v_n(x,\xi)$ with
$$
  v_0(x,\xi) = i \int_x^{+\infty} e^{2i\xi(y-x)} k(y) dy,
$$
and
\begin{equation}
  v_n(x,\xi) = \int_x^{+\infty} \int_y^{+\infty} e^{2i\xi(y-x)} k(y)^* k(z) v_{n-1}(z,\xi) dz dy.
\end{equation}
We can prove the following estimate by induction
\begin{equation} \label{Es2}
  \|v_n(x,\xi)\| \leq e^{2|Im\xi||x|} \frac{P(x,\xi)^{2n+1}}{(2n+1)!}, \quad \forall n \in \N,
\end{equation}
which implies immediately (\ref{Ml3-1}). Moreover, since $P(x,\xi)$ is uniformly bounded on $|Im\xi|<\kappa$, we deduce from (\ref{Ml3-1}) the analyticity of $M_{l3}(x,\xi)$ on the same strip. To prove (\ref{Ml3-2}), we integrate by parts in (\ref{IE1}) w.r.t the variable $y$. For all $\xi$ with $|Im\xi|<\kappa$, we obtain
\begin{eqnarray}
  M_{l3}(x,\xi) & = & -\frac{k^*(x)}{2\xi} - \frac{e^{-2i\xi x}}{2\xi} \int_x^{+\infty} e^{2i\xi y} (k^*)'(y) dy \nonumber \\ & & - \frac{k^*(x) K(x)}{2i\xi} - \frac{e^{-2i\xi x}}{2i \xi} \int_x^{+\infty} e^{2i\xi y} \Big( (k^*)'(y) K(y) - k^*(y) k(y) M_{l3}(y,\xi) \Big)  dy, \label{Es3}
\end{eqnarray}
where we have introduced the function $K(x) = \int_x^{+\infty} k(y) M_{l3}(y,\xi) dy$. Now using (\ref{AsympWNot0}) for $k$ and $k'$, (\ref{Ml3-1}) and the uniform estimate $\|K(x)\| \leq C$ for all $\xi$ with $|Im\xi|<\kappa$, we deduce that (\ref{Ml3-2}) holds when $|\xi|$ is large from (\ref{Es3}). \\ $\diamondsuit$ \\
\begin{coro} \label{al1}
  Let $\kappa$ be the positive number defined in Lemma \ref{Ml4}. Then the coefficient $a_{l1}(\xi)$ is analytic on the strip $|Im\xi|<\kappa$ and tends to $I_2$ when $|\xi| \to \infty$. Furthermore, possibly considering smaller $\kappa$ , the coefficient $a_{l1}(\xi)$ is invertible on the strip $|Im\xi|<\kappa$ and $a_{l1}^{-1}(\xi)$ is analytic and uniformly bounded there.
\end{coro}
\textit{Proof}: The first assertion is a direct consequence of (\ref{IR1}) and Lemma \ref{Ml3}. Since $a_{l2}(\xi)$ tends to $I_2$ when $|\xi| \to \infty$, $a_{l2}(\xi)$ is clearly invertible for $|\xi|$ large enough. Since $a_{l2}(\xi)$ is also invertible on the real axis (\cite{AKM}, Proposition 2.10), we conclude that $a_{l2}(\xi)$ is invertible on a strip $|Im\xi|<\e$ with $0<\e<\kappa$ small enough and that $a_{l1}^{-1}(\xi)$ is analytic and uniformly bounded on $|Im\xi|<\e$. Denoting this $\e$ by $\kappa$, we have proved the corollary. \\ $\diamondsuit$ \\

Let us put all these results together. Since $R(\xi) = -a_{l1}^{-1}(\xi) a_{l2}(\xi)$, Corollaries \ref{al2} and \ref{al1} imply that the reflection coefficient $R(\xi)$ is analytic on a strip $|Im\xi|<\kappa$ where $\kappa$ is a small enough positive number. Moreover, using the estimates of the same corollaries, we see that $\|R(.+i\eta)\| \in L^2(\R)$ for all $|\eta|<\kappa$. In fact, we have
$$
 \sup_{|\eta|<\kappa} \|R(.+i\eta)\|_{L^2} < \infty.
$$
Finally it follows from Thm IX.13 in \cite{RS} that the Fourier transform $\hat{R}(\alpha)$ satisfies the estimate
\begin{equation} \label{ExpDec}
  e^{\kappa|\alpha|} \|\hat{R}(\alpha)\| \in L^2(\R).
\end{equation}
In particular, the assumption (\ref{Cond2}) in Thm \ref{Reconstruction} is satisfied by $R(\xi)$.


We finish this paper solving the inverse problem.
\begin{theorem}
  Assume that one of the reflection matrices $L(\xi)$ or $e^{-2i\beta} R(\xi)$ appearing in (\ref{ScatMatrix}) is known on a (possibly small) interval of $\R$. Assume moreover that the mass $m$ and the charge $q \ne 0$ of the Dirac fields are known and fixed. Then the parameters $M, Q$ and $\Lambda$ of the dS-RN black hole are uniquely determined.
\end{theorem}
\textit{Proof}: We only give the proof when the reflection matrix $e^{-2i\beta} R(\xi)$ is supposed to be known on an interval $I$ of $\R$ since the proof with $L(\xi)$ can be treated the same way. We consider thus $e^{-2i\beta_1} R_1(\xi)$ and $e^{-2i\beta_2} R_2(\xi)$ two reflection matrices corresponding to parameters $M_j, Q_j$ and $\Lambda_j$ with $j=1,2$ where moreover the parameters $m, q\ne 0$ are supposed to be known and fixed. As usual we shall denote all the objects related to $e^{-2i\beta_j}R_j(\xi)$ by a lower index $j$ in what follows. Assume that $e^{-2i\beta_1} R_1(\xi) = e^{-2i\beta_2} R_2(\xi)$ for all $\xi \in I$. By analyticity, we thus have
$$
  e^{-2i\beta_1} R_1(\xi) = e^{-2i\beta_2} R_2(\xi), \quad \forall \xi \in \R.
$$
Using the procedure explained after Thm \ref{Reconstruction}, this also entails that
$$
  e^{2i\beta_1} L_1(\xi) = e^{2i\beta_2} L_2(\xi), \quad \forall \xi \in \R.
$$
Thanks to (\ref{ExpDec}) and the corresponding result for $L(\xi)$, we can apply Thm \ref{Reconstruction} (and the remarks following this Theorem). Hence we obtain the equality $e^{2i\beta_1} k_1(x) = e^{2i\beta_2} k_2(x)$ for all $x \in \R$ or equivalently
\begin{equation} \label{M0}
  e^{2i\Ga \beta_1} W_1(x) = e^{2i\Ga \beta_2} W_2(x), \quad \forall x \in \R.
\end{equation}
Now recall that $W^2$ is a positive function since
$$
W^2(x) = a_l^2(x) + b^2(x) = \Big(l+\half \Big)^2 \frac{F(r)}{r^2} + m^2 F(r),
$$
Hence taking the square of (\ref{M0}) and then the modulus, we have
\begin{equation} \label{M1}
  W_1^2(x) = a_{l,1}^2(x) + b_1^2(x) = a_{l,2}^2(x) + b_2^2(x) = W_2^2(x), \quad \forall x \in \R.
\end{equation}
Note in particular that
\begin{equation} \label{M2}
  \int_{-\infty}^{+\infty} W_1^2(s) ds = \int_{-\infty}^{+\infty} W_2^2(s) ds.
\end{equation}
Moreover by homogeneity in $l$ and since $a_l$ and $b$ are positive functions, we deduce from (\ref{M1}) that
\begin{equation} \label{M3}
  a_{l,1}(x) = a_{l,2}(x), \quad b_1(x) = b_2(x), \quad \forall x \in \R.
\end{equation}
Now since
$$
  W(x) = e^{-2i\Ga C^-(x)} (a_l(x)\Gb + b(x) \Gd),
$$
by (\ref{AntiCom}) it follows from (\ref{M0}) and (\ref{M3}) that
$$
  e^{2i\Ga \beta_1} e^{-2i\Ga C_1^-(x)} = e^{2i\Ga \beta_1} e^{-2i\Ga C_2^-(x)}, \ \quad \forall x \in \R,
$$
or equivalently that
\begin{equation} \label{M4}
  \beta_1 - C^-_1(x) = \beta_2 - C^-_2(x) + \ k\pi, \ \quad \forall x \in \R,
\end{equation}
where $k \in \Z$. Derivating (\ref{M4}), we obtain
\begin{equation} \label{M5}
  c_1(x) = c_2(x), \quad \forall x \in \R,
\end{equation}
If we let tend $x$ to $\pm \infty$, we obtain from (\ref{M5}) and (\ref{Potentials})
\begin{equation} \label{M6}
  c_{0,1} = c_{0,2}, \quad c_{+,1} = c_{+,2}.
\end{equation}

We notice eventually that (\ref{M2}) and (\ref{M6}) are precisely the conditions under which the parameters $M, Q$ and $\Lambda$ were shown to be uniquely determined in the proof of Theorem \ref{MainThm2} (see precisely the conditions (\ref{F1}) and (\ref{F2})). We thus apply the same procedure as before to end up the proof of the Theorem. $\diamondsuit$ \\


\end{document}